\documentclass[11pt]{scrartcl} 
\usepackage[a4paper, total={6in, 8in}]{geometry} 
\usepackage[singlespacing]{setspace}  

\usepackage[authoryear]{natbib}
\usepackage{amsmath, amssymb, amsthm}
\usepackage{float}
\usepackage{todonotes}
\usepackage{booktabs}
\usepackage[toc,page]{appendix}
\usepackage{graphicx}
\usepackage[ruled]{algorithm2e}

\SetKwInput{KwInput}{Input}
\SetKwInput{KwOutput}{Output}
\SetKwFor{ParFor}{for}{do in parallel}{end for}
\usepackage{svg}
\usepackage{url}
\usepackage{siunitx}
\numberwithin{equation}{section}
\usepackage{bm}

\usepackage{chngcntr}
\usepackage{multirow}
\theoremstyle{plain}

\numberwithin{equation}{section}
\renewcommand{\v}[1]{\bm{#1}}
\newcommand{\blind}{0} 

\def\*#1{\mathbf{#1}}

\usepackage{graphicx}
\usepackage{subcaption}
\usepackage{xr}

\usepackage{babel}

\title{Log predictive density score estimation in the presence of outliers}
\if0\blind
{
\author{Oskar Gustafsson$^{a}$\thanks{Corresponding author: oskar.gustafsson@stat.su.se. 
\noindent $^a$Department of Statistics, Stockholm University.} and Mattias Villani$^{a}$}
}\fi
\date{}

\begin{document}      

\maketitle

\begin{abstract}
The log predictive density score (LPDS) and related estimators of the expected log predictive density (ELPD) have become the de facto standard for Bayesian model comparison. The LPDS is a natural out-of-sample measure of predictive density performance that accounts for parameter uncertainty, is less dependent on the prior than the marginal likelihood, and is easily computed sequentially using standard Monte Carlo from posterior draws. Our article demonstrates, however, that the standard Monte Carlo estimate of the LPDS can be extremely variable and severely biased when the test data has outliers, leading to misleading conclusions and suboptimal model choices. Estimating the LPDS as the difference between the log marginal likelihood of the full dataset and the training data is shown to be a dramatically more efficient approach. This estimator is also shown to be a much more effective fallback for observations that have been flagged as problematic by the widely used PSIS-LOO model comparison framework in Stan.
\end{abstract}

\section{Introduction}
The marginal likelihood, also known as the model evidence, is fundamental for Bayesian model selection \citep{bernardo1994bayesian}, and can be viewed as a natural measure of a model's out-of-sample predictive density performance \citep{dawid1984present, geweke2010comparing, fong2020marginal, vehtari2012survey}. The marginal likelihood is, however, well known to be sensitive to the prior \citep{kass1993bayes}. This is important since it is hard to specify a detailed prior in complex hierarchical models with high-dimensional parameters commonly used in practice, and we want to prevent the comparison between models from being driven by some poorly elicited part of the prior.

The log predictive density score (LPDS) removes much of the prior sensitivity by using a portion of the data as \emph{training data} to turn the prior into a more well-behaved intermediate posterior that is subsequently used as prior for the remaining \emph{test data}. A training/test split can aim to reduce prior sensitivity and/or to estimate the generalization performance on new data. The widely used ELPD-LOO in \citet{gelman2013bayesian} uses leave-one-out cross-validation, but the actual estimators are the same as the LPDS, since the key ingredient is the same log predictive density values for observed data points; see Section \ref{sec:linreg_application} for an example using ELPD-LOO. For time series data, the training set is naturally taken as the first observations in time, with the predictive density values for each test observation computed forward in time by integrating with the posterior based on past data. For conditionally independent data without a natural order, $K$-fold cross-validation versions of the LPDS are typically used \citep{geisser1979predictive, fong2020marginal}, with ELPD-LOO as the special case with a single observation in each of the $K=n$ test sets. 

The standard approach estimates the LPDS by estimating the predictive density values for individual observations as Monte Carlo integrals over the parameter space \citep{geweke2010comparing}, where the parameter posteriors are sampled by Markov Chain Monte Carlo (MCMC) or Hamiltonian Monte Carlo (HMC); we refer to this type of estimator as the \emph{pointwise LPDS estimator}. This is easy to implement whenever a posterior sampling algorithm for the model parameters is available, and the predictive distribution conditional on the parameters is available in closed form, which is the usual case. The sum of the log of the estimated density values for individual observations is an estimate of the LPDS. This requires a full run with the posterior sampler for each observation in the test set, but the different runs can be done in parallel. For conditionally independent observations with $K$-fold cross-validated LPDS, the MCMC sampler needs to be run for each of the $K$ training folds. For leave-one-out cross-validation, $K$ equals the number of observations in the data.

The leave-one-out version of LPDS is the standard model comparison statistic in the popular Stan \citep{carpenter2017stan} framework for Bayesian inference. Stan relies on the \texttt{loo} package, which cleverly uses a single HMC run for the whole dataset followed by Pareto smoothed importance sampling (PSIS, \citet{vehtari2024pareto}) with the so-called $k$-statistic to flag observations with poorly estimated predictive density values. The predictive density for each problematic observation then needs to be re-estimated by Monte Carlo integration from a full HMC run on all the data except the problematic observation.

Our article demonstrates that the widely used LPDS estimate based on estimated log predictive density values for individual observations with parameters sampled by MCMC or HMC is problematic when there are outliers in the test data. Since an outlier is far out in the tails, the MCMC/HMC estimates are dominated by the few draws that give the largest predictive density values. This inflates the variance of the estimator for that observation and introduces a bias in the LPDS estimate, which can be large enough to alter the ranking of models. 

The LPDS idea of sacrificing part of the data to train the prior before computing the marginal likelihood on the test data has a long history in statistics, often under the name of Partial Marginal Likelihood (PML). This literature is mostly concerned with marginal likelihoods for very weak or even improper priors, and has generated several variants depending on the choice of training and test sets; see e.g. the posterior Bayes factor \citep{aitkin1991posterior}, the intrinsic Bayes factor \citep{berger1996intrinsic} and the fractional Bayes factor \citep{o1995fractional}; their asymptotic properties are studied in \citet{gelfand1994bayesian}. The PML is defined as the ratio of the marginal likelihood for the full dataset divided by the marginal likelihood of the training data; hence the log PML is a log marginal likelihood for the test data conditional on the training data, which is exactly the LPDS that the sequential approach outlined above is trying to estimate. 

The LPDS can therefore alternatively be estimated from the logarithm of the PML, i.e. as the difference between the log marginal likelihood estimate for the full data and the log marginal likelihood estimate for the training data; we refer to this type of estimator as the \emph{PML estimator}. This simple alternative way to estimate the LPDS is rarely used in the literature; an exception is Siddhartha Chib who uses a related approach in a series of recent papers and software (see e.g.~\citet{chib2026testing, chib2025structural}), where the posterior based on a training set with $15\%$ of the data is used to set the hyperparameter values for the prior used for the remaining test observations. A main contribution of our article is to show that the PML estimator of the LPDS can be vastly more efficient than the commonly used pointwise LPDS estimator, particularly when the test data have outliers. \citet{o1995fractional} discusses outliers, but only in arguing that the fractional marginal likelihood is more robust to outliers than alternative PMLs when the training sample is minimal, which is the focus of that strand of literature. 

The success of the PML estimator hinges on having an accurate marginal likelihood estimator, which is not always available. However, we show empirically that the PML estimator based on Chib's marginal likelihood estimator \citep{chib1995marginal} for models amenable to Gibbs sampling, or the Bridge sampling estimator \citep{meng1996simulating} for more general models give very accurate LPDS estimates when used in the PML estimator. Alternatively, the Laplace approximation of the marginal likelihood \citep{kass1993bayes} gives a fast optimization-based estimate which is often accurate, unless the training sample is small.

The PML estimator can also be used to assess the predictive density score for selected subperiods of the data, or even to compute the one-step-ahead predictive density score for every observation in the test sample. In particular, we show that the PML estimator is a more efficient way of computing the leave-one-out predictive density for the individual observations flagged by the $k$-statistic in PSIS compared to the recommended MCMC/HMC fallback.

\section{The pointwise LPDS estimator and outliers} \label{sec:theory}

This section considers a time series setup, but the ideas apply equally well to conditionally independent data or other dependence structures.

\subsection{The marginal likelihood and log predictive density score}

The marginal likelihood for a given model is the joint density of the observed data
\begin{equation}
    p(\bm{y}_{1:T}) = \int p(\bm{y}_{1:T} \vert \v\theta)p(\v\theta)d\v\theta,
\end{equation}
where $\bm{y}_{1:T}=(\v y_1,\ldots,\v y_T)$ is a potentially vector-valued time series, and $p(\v\theta)$ is the prior distribution for the model parameter vector $\v\theta$. The log marginal likelihood can be decomposed into a sequence of one-step-ahead predictive densities
\begin{equation}
\log p(\bm{y}_{1:T}) = \sum_{t=1}^{T} \log p(\bm{y}_{t} | \bm{y}_{1:t-1}), \label{eq:mlChain}
\end{equation}
where the first term at $t=1$ is $p(\bm{y}_{1})$. 

The marginal likelihood is, however, known to be sensitive to the choice of prior \citep{kass1993bayes}, which in a time series context has the largest effect on the terms $p(\bm{y}_{t} | \bm{y}_{1:t-1})$ for the first observations in the time series. The Log Predictive Density Score (LPDS) therefore uses the first $t_0$ observations to train the prior into a more well-behaved intermediate posterior and computes the marginal likelihood for the remaining observations using this new prior. This corresponds to discarding the first $t_0$ terms in \eqref{eq:mlChain}
\begin{equation}
    \mathrm{LPDS}_{t_0+1:T} = \sum_{t=t_0+1}^{T} \log p(\bm{y}_{t} | \bm{y}_{1:t-1}). \label{eq:LPDSdef}
\end{equation}
The individual terms in the sum $p(\bm{y}_{t} | \bm{y}_{1:t-1})$ are called \emph{LPDS contributions}.

\subsection{Variance and bias of the pointwise LPDS estimator}

The standard way to estimate the LPDS contribution at time $t$,
\begin{equation}
    \text{LPDS}_t = \log p(\v y_t \mid \v y_{1:t-1}) = \log \int p(\v y_t \mid \v \theta) p(\v \theta \mid \v y_{1:t-1}) d\v\theta.
\end{equation}
is by the log of the Monte Carlo estimate
\begin{equation}
    \widehat{p}(\v y_t \mid \v y_{1:t-1}) = \frac{1}{M} \sum_{m=1}^M p(\v y_t \mid \v\theta^{(m)}, \v y_{1:t-1}), \label{eq:MCestimator}
\end{equation}
where $\v \theta^{(m)}$ are $M$ draws from the previous posterior, $\v \theta^{(m)} \sim p(\v\theta \mid \v y_{1:t-1})$.
The estimator $\widehat{p}_t := \widehat{p}(\v y_t \mid \v y_{1:t-1})$ is unbiased for the predictive density evaluation, $p_t := p(\v y_t \mid \v y_{1:t-1})$. However, when $\v y_t$ is far out in the tail, the Monte Carlo estimate will be dominated by a few draws and have a large variance.

Although $\widehat{p}_t$ is an unbiased estimator of the density value $p_t$, the $\log(\widehat{p}_t)$ terms in the LPDS estimator will be biased downward, $\mathbb{E}( \log \widehat{p}_t ) \leq \log p_t$ for any finite $M$, due to the concavity of the logarithmic function and Jensen's inequality. Lemma 2 in \citet{pitt2012some} can be used to quantify the bias. Translated to the current setting, their lemma shows that if $\widehat{p}_t$ is unbiased and satisfies a central limit theorem, then, informally stated, for large $M$
\begin{equation}\log \widehat{p}_t \overset{\mathrm{approx}}{\sim}N\bigg(\log p_t - \frac{\mathrm{Var}( \widehat{p}_t)}{2p_t^2}, \frac{\mathrm{Var}(\widehat{p}_t)}{p_t^2} \bigg).\label{eq:CLT_logphat}
\end{equation}
See \citet{pitt2012some} and \citet{berard2014lognormal} for the formal convergence result and proof. This result is in particular applicable when the estimator in \eqref{eq:MCestimator} is computed from Monte Carlo or MCMC draws satisfying a central limit theorem. The asymptotic variance of $\widehat{p}_t$ is then  $\mathrm{Var}(\widehat{p}_t) \approx \frac{1}{M_{\mathrm{eff}}}\mathrm{Var}_{\v\theta \vert y_{1:t-1}}\big(p(\v y_t \mid \v\theta, \v y_{1:t-1})\big)$, where $M_{\text{eff}}$ is the effective sample size, taking into account the MCMC dependence in the $\v \theta$ draws.

The result in \eqref{eq:CLT_logphat} shows that 
\begin{equation}
    \mathrm{Bias}(\log \widehat{p}_t) \approx - \frac{\mathrm{Var}( \widehat{p}_t)}{2p_t^2}.
    \label{eq:bias_general}
\end{equation}
The bias is therefore large when $\mathrm{Var}(\widehat{p}_t)$ is large and/or the marginal likelihood contribution $p_t$ is small.
When the observation $\v y_t$ is an outlier, the true predictive density, $p_t$, is very small, and the MCMC draws of $\v \theta$ yield highly skewed density values. This causes the estimator's variance to be high relative to its squared mean ($p_t^2$), which makes the pointwise LPDS estimator biased and unstable. Note that this is separate from the bias discussed by \citet{vehtari2017practical} coming from the in-sample optimism or from estimating the generalization performance from cross-validation with training samples that are smaller than the actual dataset used to predict new observations.

To illustrate the bias of the pointwise LPDS estimator, we consider a simple $K$-dimensional iid multivariate normal model $\v y_t \overset{\mathrm{iid}}{\sim} \mathcal{N}_K(\v \mu, \v\Sigma)$ with a known covariance matrix $\v\Sigma$. We use a standard conjugate normal prior $\v\mu \sim \mathcal{N}_K(\v\mu_0, \frac{1}{\kappa_0}\v\Sigma)$, where $\kappa_0 > 0$ acts as a prior sample size. \cite{sivula2025uncertainty} recently used a similar simplified analytical framework to study the sampling variability of LOO-CV estimators across dataset. Our focus is strictly on the Monte Carlo approximation error for a given fixed dataset. 

Given the data up to observation $t-1$, standard Bayesian updating yields the posterior distribution 
\begin{equation}
    \v\mu \mid \v y_{1:t-1} \sim \mathcal{N}_K(\hat{\v \mu}_{t-1}, c_{t-1}\v \Sigma),
\end{equation}
where $\hat{\v\mu}_{t-1}= \frac{\kappa_0}{\kappa_0 + t-1} \v\mu_0 + \frac{t-1}{\kappa_0 + t -1 } \bar{\v y}_{t-1}= c_{t-1}(\kappa_0 \v\mu_0 + (t-1)\bar{\v y}_{t-1})$ is the posterior mean and $c_{t-1} = \frac{1}{\kappa_0 + t - 1}$. The scalar $0 < c_{t-1} < 1$ scales the parameter uncertainty. We wish to evaluate the predictive density for a new, potentially extreme observation $\v y_{t}$.

The Monte Carlo estimator for the predictive density $p(\v y_{t} \mid \v y_{1:t-1})$ is given by:
\begin{equation}
    \widehat{p}_t = \frac{1}{M} \sum_{m=1}^M \mathcal{N}(\v y_{t} \mid \v\mu^{(m)}, \v\Sigma), \quad \v\mu^{(m)} \sim p(\v\mu \mid \v y_{1:t-1}).
\end{equation}
Each term in the sum is proportional to $\exp(-\frac{1}{2} Q_t^{(m)})$, where $Q_t^{(m)} = (\v y_{t} - \v \mu^{(m)})^T \v \Sigma^{-1} (\v y_{t} - \v\mu^{(m)})$. The standardized distance $\hat{p}_t\propto Q_t^{(m)}/c_{t-1}$ therefore follows a non-central $\chi^2(K)$ distribution with non-centrality parameter
\begin{equation}
    \lambda_t=\frac{1}{c_{t-1}}D_t^2
\end{equation}
where $D_t^2=(\v y_{t}-\hat{\v \mu}_{t-1})^T\v\Sigma^{-1}(\v y_{t}-\hat{\v \mu}_{t-1})$ is the squared Mahalanobis distance of the new data point to the posterior mean based on the previous $t-1$ observations. Using the moment generating function of the non-central $\chi^2$ we can compute the variance of $\hat{p}_t$ and plug it into \eqref{eq:bias_general} to obtain
\begin{equation} \label{eq:bias}
    \mathrm{Bias}(\log \hat{p}_t)\approx - \frac{1}{2M_{\text{eff}}} \left[ \left( 1 + \frac{c_{t-1}^2}{1+2c_{t-1} } \right)^{K/2} \exp\left( \frac{c_{t-1} D_t^2}{(1+c_{t-1})(1+2c_{t-1})} \right) - 1 \right].
\end{equation}
See Appendix \ref{sec:deriveBias} for the proof. 

Since $\mathrm{Var}(\log \hat{p}_t) \approx -2\cdot\mathrm{Bias}(\log \hat{p}_t)$, the following comments on the bias are equally valid for the variance. First, the bias in \eqref{eq:bias} vanishes as the posterior uncertainty $c_{t-1} = 1/(\kappa_0 + t -1)$ goes to zero, i.e.~as the sample size and/or prior precision increases. Importantly, the bias grows exponentially with the squared Mahalanobis distance $D_t^2$. For small values of $c_{t-1}$, the argument of the exponential function is $c_{t-1} D_t^2$ to first order, so the effect of an outlier is mitigated as $c_{t-1}$ decreases. Since the expected value of $D_t^2$ scales linearly with the dimension $K$, the bias from an outlier tends to be worse in high-dimensional models. Finally, there is a bias even for the most typical observation where $D_t=0$, and this baseline bias is easily seen to be negative.

Section \ref{sec:deriveBias} in the Supplementary Material uses simulation to show that the bias approximation in \eqref{eq:bias} is quite accurate for outliers, both in $K=1$ and $K=20$ dimensions.

\subsection{Illustrative example with a multivariate normal model}\label{sec:normalunknown}

We now consider the more realistic scenario where both the mean vector $\v \mu$ and the covariance matrix $\v \Sigma$ are unknown and jointly estimated. We use the conjugate Normal-Inverse-Wishart (NIW) prior:
\begin{align}
    \v\Sigma &\sim \mathcal{IW}(\nu_0, \v\Lambda_0) \\
    \v\mu \mid \v\Sigma &\sim \mathcal{N}_K(\v\mu_0, \v\Sigma / \kappa_0)
\end{align}
where $\nu_0$ is the prior degrees of freedom for $\v\Sigma$, and $\v\Lambda_0$ is the prior scale matrix.

Given a sequence of observations up to time $t-1$ the joint posterior $p(\v\mu, \v\Sigma \mid \v y_{1:t-1})$ is a NIW distribution with updated parameters $\v\mu_{t-1}, \kappa_{t-1}, \nu_{t-1} \text{ and } \v\Lambda_{t-1}$. The  predictive distribution for $\v y_{t}$ is obtained by integrating with respect to the sequential posterior density $p(\v\mu,\v\Sigma \vert \v y_{1:t-1})$ and is known to be a $K$-dimensional multivariate Student-t distribution \citep[see e.g.][sec.~3.6]{gelman2013bayesian}:
\begin{equation}
    \v y_{t} \mid \v y_{1:t-1} \sim t_{\nu_{t-1} - K + 1}\left( \v\mu_{t-1}, \, \frac{\kappa_{t-1} + 1}{\kappa_{t-1}(\nu_{t-1} - K + 1)} \v\Lambda_{t-1} \right).
    \label{eq:exact_predictive_integral}
\end{equation}

We will now assume that the true predictive density is not known and needs to be estimated. The pointwise LPDS estimator approximates the integrals with respect to $\v\mu$ and $\v\Sigma$ using $M$ posterior draws $\v\mu^{(m)},\v\Sigma^{(m)} \sim p(\v\mu,\v\Sigma \vert \v y_{1:t-1})$
\begin{equation}
    \hat{p}(\v y_{t} \mid \v y_{1:t-1}) = \frac{1}{M} \sum_{m=1}^M \mathcal{N}_K\left(\v y_{t} \mid \v\mu^{(m)}, \v\Sigma^{(m)}\right).
\end{equation}
This implies that we are attempting to approximate a heavy-tailed multivariate Student-t distribution with a finite mixture of multivariate normal components \citep{bitto2019achieving}. 

\begin{figure}[h!]
    \centering
    \includegraphics[width=0.9\linewidth]{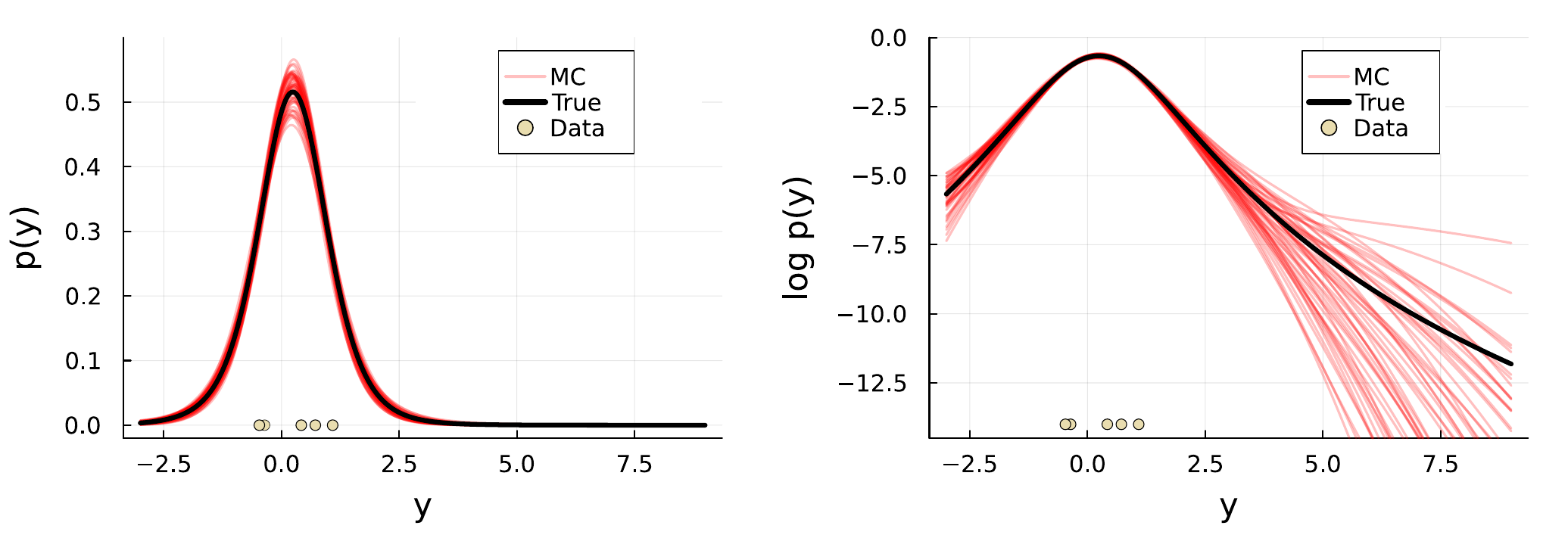} 
    \caption{Estimates of the predictive densities for the univariate normal model with unknown mean and variance on original scale (left) and log scale (right). Each red line is an estimate from one of $50$ independent Monte Carlo runs, each with 100 draws. The beige dots are the five data points in the training sample.}
    \label{fig:spaghettiPlot}
\end{figure}

To build visual intuition for why this approximation fails in the tails, Figure \ref{fig:spaghettiPlot} illustrates the predictive densities implied by 50 independent Monte Carlo runs, each with 100 draws, for the univariate case with $K=1$. We use only $T=5$ simulated data points (beige points) to mimic the large parameter uncertainty that a much larger sample size would give in a complex high-dimensional model. 

The graph on the left of Figure \ref{fig:spaghettiPlot} shows that the true predictive distribution is seemingly well captured by the Monte Carlo estimates. However, in the right graph, we inspect the same predictive distributions on the log-scale on which LPDS is evaluated. A normal density decays quadratically on the log-scale, whereas the Student-t distribution has polynomial tails that decay much more slowly. When we use a finite posterior sample, we typically lack the extreme parameter draws required to represent the heaviest tails. Without these components, the finite mixture of normals decays too rapidly towards negative infinity, which can create a severe downward bias whenever $y_{t}$ is an outlier.

Although the problem is clearly visible in one dimension, it gets much worse as the dimensionality $K$ increases, as we will see later. A finite set of posterior draws becomes increasingly sparse in high-dimensional parameter spaces, and it becomes even less likely to sample the extreme $\v\mu$ and $\v\Sigma$ parameters needed to cover a multivariate outlier. 

For this model, we can easily improve the Monte Carlo performance by analytically integrating out the variance parameter $\Sigma$ forming a Monte Carlo average only over the $\mu$ draws. The resulting mixture of Student-t components fits the tails much better, as illustrated in the next section. A similar point is demonstrated in \citet{bitto2019achieving}, showing that the pointwise LPDS estimator has better performance in a state-space model when the state is integrated out with the Kalman filter. Nevertheless, far from all models allow for such marginalization, and the next section shows that an LPDS estimator from two marginal likelihood estimates is a much better alternative in the presence of outliers.

\section{Estimating the LPDS via marginal likelihoods}

The pointwise LPDS estimate dominates in the more recent literature and statistical practice. It is simple to code and is directly applicable for any model where we can generate posterior draws and evaluate the model distribution conditional on the parameters. A drawback is that it requires $T - t_0$ separate Markov Chain Monte Carlo (MCMC) runs, which is time-consuming. This computational cost can be reduced by parallelization and importance sampling techniques \citep{vehtari2024pareto}, but a serious drawback of the pointwise LPDS estimator is, as we show in this article, that it is biased and highly variable in the presence of outliers. This section proposes an alternative estimator which is much more robust to outliers.

\subsection{The PML estimator}

The problems with the pointwise LPDS estimate can be avoided with an alternative estimator based on the chain rule decomposition of the marginal likelihood in \eqref{eq:mlChain}. By simply rearranging the identity, the LPDS can be isolated as the difference of two log marginal likelihoods
\begin{equation}
\mathrm{LPDS}_{t_0+1:T} = \log p(\bm{\v y}_{1:T}) - \log p({\v y}_{1:t_0});
\end{equation}
The LPDS can therefore be estimated from a difference between the log marginal likelihood of the full sample and the log marginal likelihood of the training sample. This is the partial marginal likelihood (PML) used in the mid 90's literature on model comparison with improper priors (see e.g. \citet{aitkin1991posterior, berger1996intrinsic, o1995fractional}). However, our motivation for using it here for estimating the LPDS is different: it is fast to compute, and it is much more robust to outliers.

The PML estimator of the LPDS requires a precise estimator of the marginal likelihood. We demonstrate here that the PML estimator with either Chib's method \citep{chib1995marginal} or bridge sampling \citep{meng1996simulating} is much more efficient and less biased than the pointwise LPDS estimator.

A main difference between the pointwise estimator and the PML estimator is that density evaluations of \emph{individual} observations appear in the pointwise estimator, whereas the PML estimator typically involves only \emph{average} summary statistics of the full sample and the training sample; this makes the PML estimator more robust to outliers. The most direct way to see this is in the exponential family class of models, where the distribution of the multivariate $\v y$ is of the form \citep{bernardo1994bayesian}
\begin{equation}
  p(\v y\vert \v\theta)=h(\v y)\exp\Big(\eta(\v\theta)^\top \bm{t}(\v y)-A(\v\theta)\Big),
\end{equation}
for a vector $\bm{t}(\v y)$ of sufficient statistics. Let $p(\v\theta)$ be the prior distribution for $\v\theta$. The marginal likelihood is given by
\begin{equation}
    p(\v y_1,\ldots, \v y_n)=\prod_{i=1}^n h(\v y_i) \int \exp\Big(n \v\eta(\v\theta)^\top \bar{\bm{t}} - n A(\v\theta)\Big) p(\v\theta) d\theta,
\end{equation}
where $\bar{\bm{t}} = n^{-1}\sum_{i=1}^n \bm{t}(\v y_i)$ is the average of the sufficient statistics. Any Monte Carlo integral approximation of the above integral will only be a function of the data through the average of the sufficient statistics $\bar{\bm{t}}$. The factor $\prod_{i=1}^n h(\v y_i)$ can depend on individual data points, but it is not a function of the parameters $\v\theta$, and thus does not contribute to the variance of the Monte Carlo approximation.

Another view of the stability of the PML estimator is by inspecting specific estimators of the marginal likelihood. For example, for models estimated by Gibbs sampling, \citet{chib1995marginal} uses an identity for the marginal likelihood obtained by inverting Bayes' theorem
\begin{equation}
p(\v y_{1:t}) = \frac{p(\v y_{1:t} \vert \v\theta)p(\v\theta)}{p(\v\theta \vert \v y_{1:t})},
\end{equation}
which holds for any $\v\theta$. By evaluating the identity for both $p(\v y_{1:t})$ and $p(\v y_{1:t-1})$ at the same fixed parameter point $\v\theta^*$, the out-of-sample predictive density at time $t$ can be expressed as
\begin{equation} \label{eq:ratioChib}
    p(\v y_t \mid \v y_{1:t-1}) = p(\v y_t \mid \v\theta^*, \v y_{1:t-1}) \frac{p(\v\theta^* \mid \v y_{1:t-1})}{p(\v\theta^* \mid \v y_{1:t})}.
\end{equation}
This representation isolates the sources of estimation error. The new observation, a potential outlier, $\v y_t$ is evaluated only in the likelihood contribution, $p(\v y_t \mid \v\theta^*, \v y_{1:t-1})$, which is computed analytically and is therefore completely free of Monte Carlo variability. This shows that the MCMC approximation is restricted entirely to the ratio of the two posterior ordinates. By the recommendation in \citet{chib1995marginal} to choose the anchor point $\v\theta^*$ in a high-density region (e.g. the posterior mode), these ordinates are estimated with high precision. We note that (\ref{eq:ratioChib}) suggests that we only need to compute the likelihood for the single point, and the two posterior densities at $\v\theta^*$. However, the extra computations of evaluating the prior and the remaining likelihood terms are negligible, and we use the full marginal likelihoods in our examples.

For more general models, including those estimated via Hamiltonian Monte Carlo (HMC), bridge sampling \citep{meng1996simulating} achieves a similar stability by replacing the single anchor point with a continuous proposal distribution, $g(\v\theta)$. The normalizing constant $p(\v y_{1:t})$ is computed via the expected ratio between the unnormalized posterior $p^*(\v\theta \mid \v y_{1:t})$ and $g(\v\theta)$. Crucially, $g(\v\theta)$ is explicitly fitted to the moments of the MCMC draws from $p(\v\theta \mid \v y_{1:t})$, and is thus centered in the same region of the parameter space. The bulk of the Monte Carlo expectation is evaluated where both distributions have their mass.


\subsection{Revisiting the normal example} \label{sec:comparison}

We here revisit the univariate simulation example in Section \ref{sec:normalunknown} with a normal model with an unknown variance. The LPDS was shown to be tractable with the predictive densities as $t$-distributions. We use the same data settings and priors as in that example. The following four LPDS estimators are compared in the presence of an outlier:
\begin{itemize}
    \item \textit{Pointwise LPDS:} The standard LPDS approach, approximating the predictive density as a finite mixture of normal distributions using MCMC draws from the previous period's posterior.
    \item \textit{Marginalized pointwise LPDS:} The analytically reduced estimator, where the variance parameter is integrated out, resulting in a finite mixture of Student-$t$ components.
    \item \textit{PML with bridge sampling:} The PML estimator with marginal likelihoods estimated with bridge sampling.
    \item \textit{PML with Chib's estimator:} The PML estimator with marginal likelihoods estimated with Chib's estimator.
\end{itemize}

\begin{figure}
    \centering
    \includegraphics[width=0.9\linewidth]{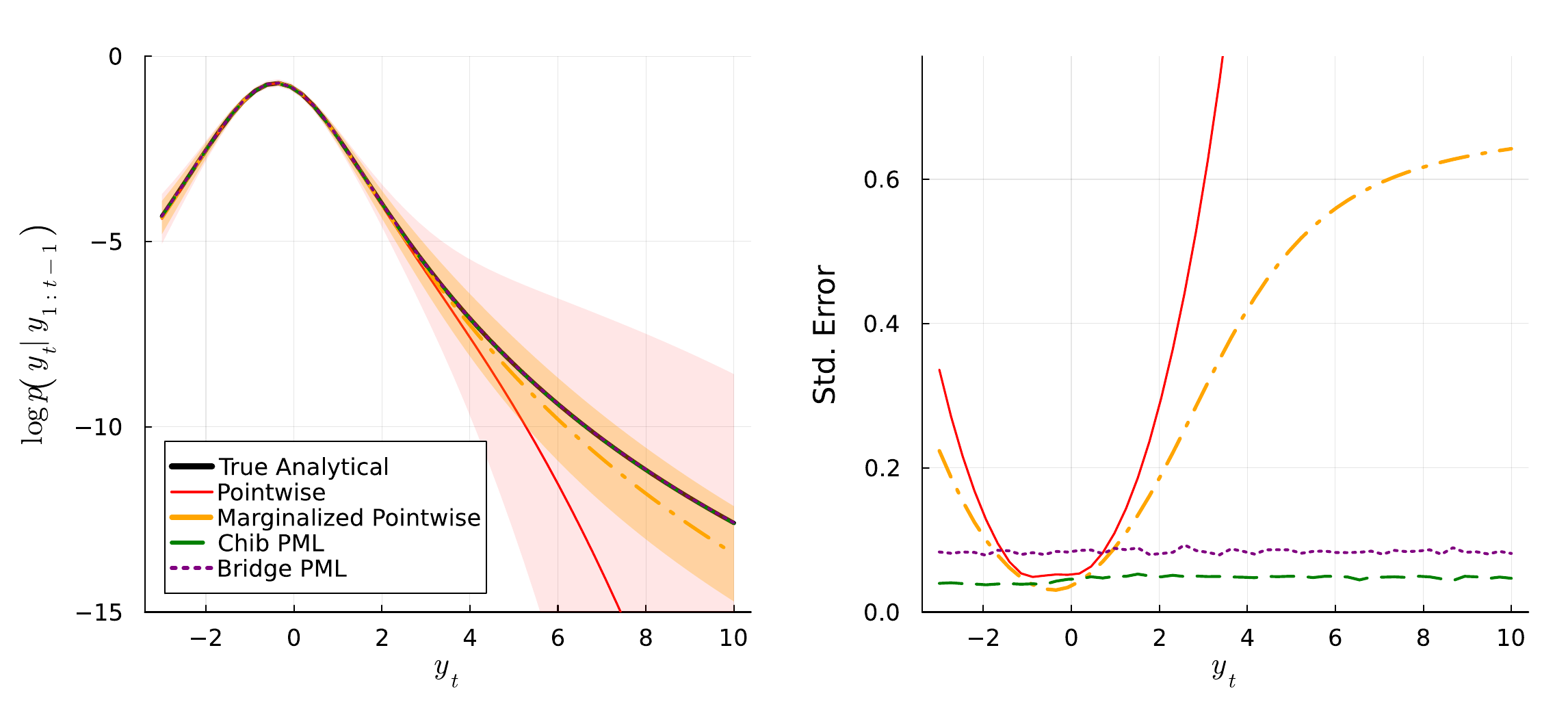}
    \caption{Estimated predictive density values as a function of the observation $y_t$ for the four compared estimators (left). The true predictive density value is the black line and the shaded regions are 2 std.~error bands for the pointwise LPDS estimator without marginalization (red bands) and with marginalization (orange bands). The error bands for the two PML estimators are very tight and are excluded here for clarity. The right hand graph displays the standard errors of all four estimators as a function of the observation $y_t$.}
    \label{fig:UnivariateComparison}
\end{figure}

All methods are based on the same output from a Gibbs sampler that alternates between sampling the unknown $\mu$ and $\sigma^2$, pretending that the closed-form solution is unavailable. The iterative bridge sampler uses a normal proposal for $\mu$ and a log-normal for $\sigma^2$, both at the mean and variance of the posterior draws.
Note that although we can marginalize out $\sigma^2$ for both Chib and the Bridge sampling estimators, this is less interesting for our comparisons.

Figure \ref{fig:UnivariateComparison} shows that marginalization gives a substantial improvement for the pointwise LPDS estimator when the outlier is severe, both in reducing the bias and for lowering the standard errors. The most striking feature of Figure \ref{fig:UnivariateComparison}, however, is the excellent performance of both PML estimators, virtually without bias even at the most extreme outliers and standard errors that are flat over the outlier space. The PML estimator using Chib's estimator is more precise, as expected, since it uses more information from the problem setting. The advantage of bridge sampling is that it does not need the full conditionals, and is therefore applicable for any model where we can sample from the posterior and evaluate the unnormalized posterior. 

\subsection{A trimmed PML estimator}\label{sec:trimmedPML}

It is not uncommon to evaluate predictive performance while excluding specific extreme outliers or periods (e.g., the COVID pandemic) that may otherwise disproportionately distort model selection. This is easily accommodated with the pointwise LPDS estimator since it estimates the LPDS contributions for every time point and any individual observation or period can therefore be subtracted from the LPDS.

The PML estimator can also be used to estimate an individual LPDS contribution at some time $t$, or more generally whole time stretches $t,t+1.\ldots,t+\ell$, by a difference of two marginal likelihood estimates
\begin{equation}
   \mathrm{LPDS}_{t:t+\ell} = \log p(\v y_{t:t+\ell} \vert \v y_{1:t-1} ) = \log p(\v y_{1:t+\ell} ) - \log p(\v y_{1:t-1} ).
\end{equation}
This property implies that we can selectively exclude a set of specific outlier periods from the evaluation. Let $\mathcal{K}=\{\mathcal{K}_1,\ldots,\mathcal{K}_S\}$ denote the set of $S$ non-overlapping time segments $\mathcal{K}_s = t_s:t_s + \ell_s$ that we want to remove from the evaluation. A trimmed LPDS simply subtracts the LPDS contributions of the removed segments
\begin{equation}
\begin{aligned}
    \mathrm{LPDS}_{\text{trim}} = \mathrm{LPDS}_{t_0:T} -  \sum_{s = 1}^S \mathrm{LPDS}_{\mathcal{K}_s},
\end{aligned}\label{eq:trimmedLPDS}
\end{equation}
where $\mathrm{LPDS}_{\mathcal{K}_s}$ is the log predictive score for segment $\mathcal{K}_s$.
This trimmed evaluation only requires estimating the marginal likelihood at the endpoints ($T$ and $t_0$) and at the boundaries of the unwanted intervals at a computational cost of $O(S)$. Note that $K$-fold cross-validated LPDS for conditionally independent data can be efficiently implemented using \eqref{eq:trimmedLPDS} by letting each fold be a segment.

Finally, the leave-one-out cross-validation approach with Pareto Smoothed Importance Sampling (PSIS) \citep{vehtari2017practical} is the default model comparison approach in probabilistic programming frameworks such as Stan \citep{carpenter2017stan} and Turing.jl \citep{ge2018turing}. PSIS is very efficient, but can fail for high-leverage outliers. The above mentioned frameworks will flag such observations and the recommended approach is to re-estimate their LPDS contributions using the pointwise LPDS estimator. We show in Section \ref{sec:linreg_application} that the bias and variability of this fallback can lead to erroneous model selection and that the PML estimator is a much better fallback estimator.

\section{Hyperparameter selection in vector autoregressive model for US macroeconomic data}\label{sec:BVARapplication}

This section contrasts the performance of the pointwise and PML estimator of the LPDS in a multivariate time series model widely used for forecasting and policy analysis in macroeconomics. There has been recent interest in selecting the hyperparameters in the so called Minnesota prior \citep{doan1984forecasting}; see e.g.~\citep{carriero2012forecasting, giannone2015prior, gustafsson2023bayesian}.

\subsection{Model, data and priors}

The steady-state Bayesian vector autoregression (BVAR) \citep{villani2009steady} is of the form
\begin{equation}
    \v\Pi(L)(\v y_t-\v \psi)=\v \varepsilon_t, \qquad \v\varepsilon_t\sim \mathcal{N}(\v 0, \v\Sigma),
\end{equation}
where $\v\Pi(L)=\v I- \v\Pi_1 L - \ldots - \v\Pi_p L^p$ is the lag polynomial and $\v\Pi_k$ are the matrices of autoregressive coefficients, $\v\psi$ is a vector of unconditional means  (steady-states), and $\v \Sigma$ is a positive definite covariance matrix. The process is assumed to be stationary by the usual condition on the roots of the lag polynomial. Section \ref{subsec:t-errors} explores a more robust version of the model with student-$t$ disturbances.

The data consist of 243 quarterly observations on Real GDP, GDP deflator, and the FED funds rate. The data is restricted to the period 1959Q2-2019Q4, to avoid extreme movements due to the pandemic. The first two series are transformed to stationarity by computing $400\times$ $\log$ quarterly changes, while the FED funds rate is untransformed. 
The model is estimated using $p=4$ lags.

We use a standard Minnesota prior with hyperparameters for  \textit{overall shrinkage} $\lambda_1$, \textit{cross equation shrinkage} $\lambda_2$ and \textit{lag decay shrinkage} $\lambda_3$. The remaining priors are:
\begin{align}
    \v\psi  &\sim \mathcal{N}\left((3,2,5)^\top, 0.25^2\v I \right),  \\
    \v\Sigma &\sim \mathcal{IW}(5, \mathrm{Diag}(s^2_{\mathrm{ols},40})),\\
    \mathrm{vec}(\v\Pi) &\sim \mathcal{N}(\underline{\v\pi},\underline{\v\Omega})  
\end{align}
where $\mathrm{Diag}(s^2_{\mathrm{ols},40})$ is a diagonal matrix containing the residual variances of univariate AR($p$) models, for the $40$ quarters of data, for each series separately. The vector $\mathrm{vec}(\v\Pi)$ is multivariate normal with mean of the first own lag being $(0,0.6,0.6)^\top$, the rest of the mean coefficients are zero, and the covariance matrix is given by the Minnesota prior with the hyperparameters given above.

\subsection{Results from a model with normal errors}
To keep things simple for illustration we restrict the hyperparameter search to the overall shrinkage hyperparameter $\lambda_1$, for fixed $p=4, \lambda_2=0.5$, and $\lambda_3=1$. The LPDS is estimated for a grid of $\lambda_1$ values using the pointwise and PML estimators, and the two LPDS-curves are compared to see the effect on the choice of hyperparameter.

\begin{figure}[htbp]
    \centering
    
    \begin{minipage}[c]{0.6\textwidth}
        \centering
        \includegraphics[width=0.9\linewidth]{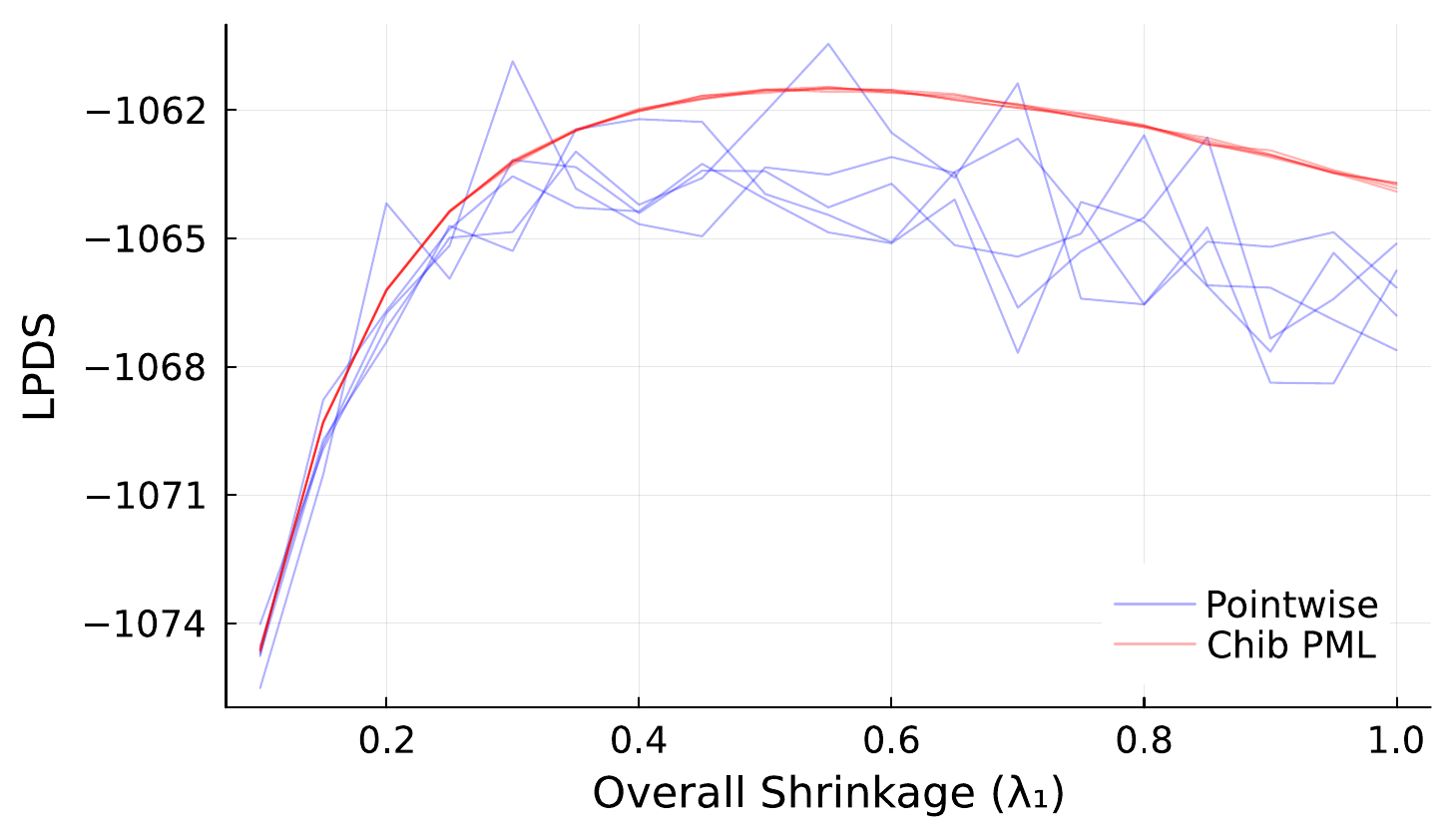}
    \end{minipage}
    \hfill
    \begin{minipage}[c]{0.39\textwidth}
    \vspace{-0.1cm}
        \centering
    \begin{tabular}{crr}
        \toprule
        Run & Pointwise & Chib PML \\
        \midrule
        1 & 0.40 & 0.50 \\
        2 & 0.80 & 0.55 \\
        3 & 0.55 & 0.55 \\
        4 & 0.30 & 0.55 \\
        5 & 0.45 & 0.55 \\
        \bottomrule
    \end{tabular}
\small{Selected overall shrinkage level}  \hspace{0.5cm}
     \end{minipage}
    
    \vspace{4mm}
    
    \caption{Comparison of LPDS from 1970Q1 to 2019Q4 using $5000$ MCMC draws with a burn-in of $1000$  for five random runs over a grid of overall shrinkage values ($\lambda_1$). \textit{Left:} illustration of the variability of the LPDS estimates over different runs. \textit{Right:} selected degree of shrinkage based on the LPDS estimate.}
    \label{fig:lambdaGrid}
\end{figure}

We use the first $43$ quarters (1959Q2-1969Q4) as training data and the subsequent $200$ observations in 1970Q1-2019Q4 for evaluating the LPDS. The pointwise LPDS is estimated by running the Gibbs sampler in \cite{villani2009steady} for every evaluation point in the test data, with each Gibbs run consisting of $5000$ draws after a burn-in of $1000$ draws; hence, this requires $200$ full Gibbs runs with increasing time series lengths. The PML estimator, on the other hand, requires just two marginal likelihood estimates: one for the full data, up to 2019Q4, and, the other for the data up to 1969Q4. Both estimators are repeated in five independent runs to get a sense of the variability. 

\begin{figure}
    \centering
    \includegraphics[width=0.8\linewidth]{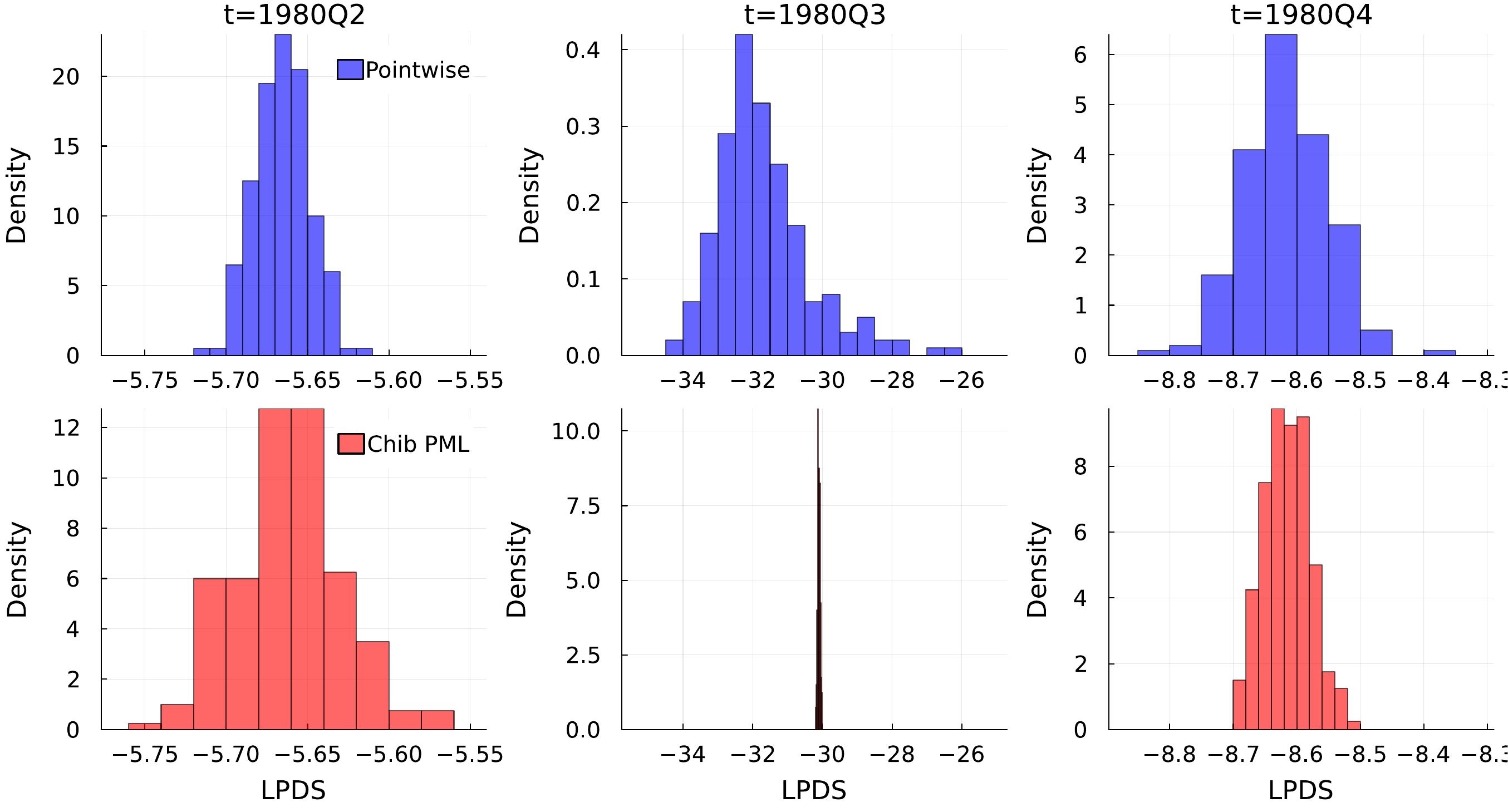}
    \caption{Bayesian VAR with normal error terms. Distribution of the estimates of the LPDS contributions for the three quarters during the Volcker shock period over $200$ runs with different seeds. Top row plots the pointwise LPDS, while the bottom row plots the PML based on Chib's estimator. Each estimate is based on $5000$ Gibbs sampling draws, where the first 1000 were discarded as a burn-in.}
    \label{fig:CompLPDSt}
\end{figure}

Figure \ref{fig:lambdaGrid} clearly shows that the pointwise LPDS estimator is extremely noisy and unreliable here; it suggests widely different optimal hyperparameter values over the different runs. The PML estimator, on the other hand, is so stable across runs that the LPDS curves are almost indistinguishable. 

\begin{figure}
    \centering
    \includegraphics[width=0.8\linewidth]{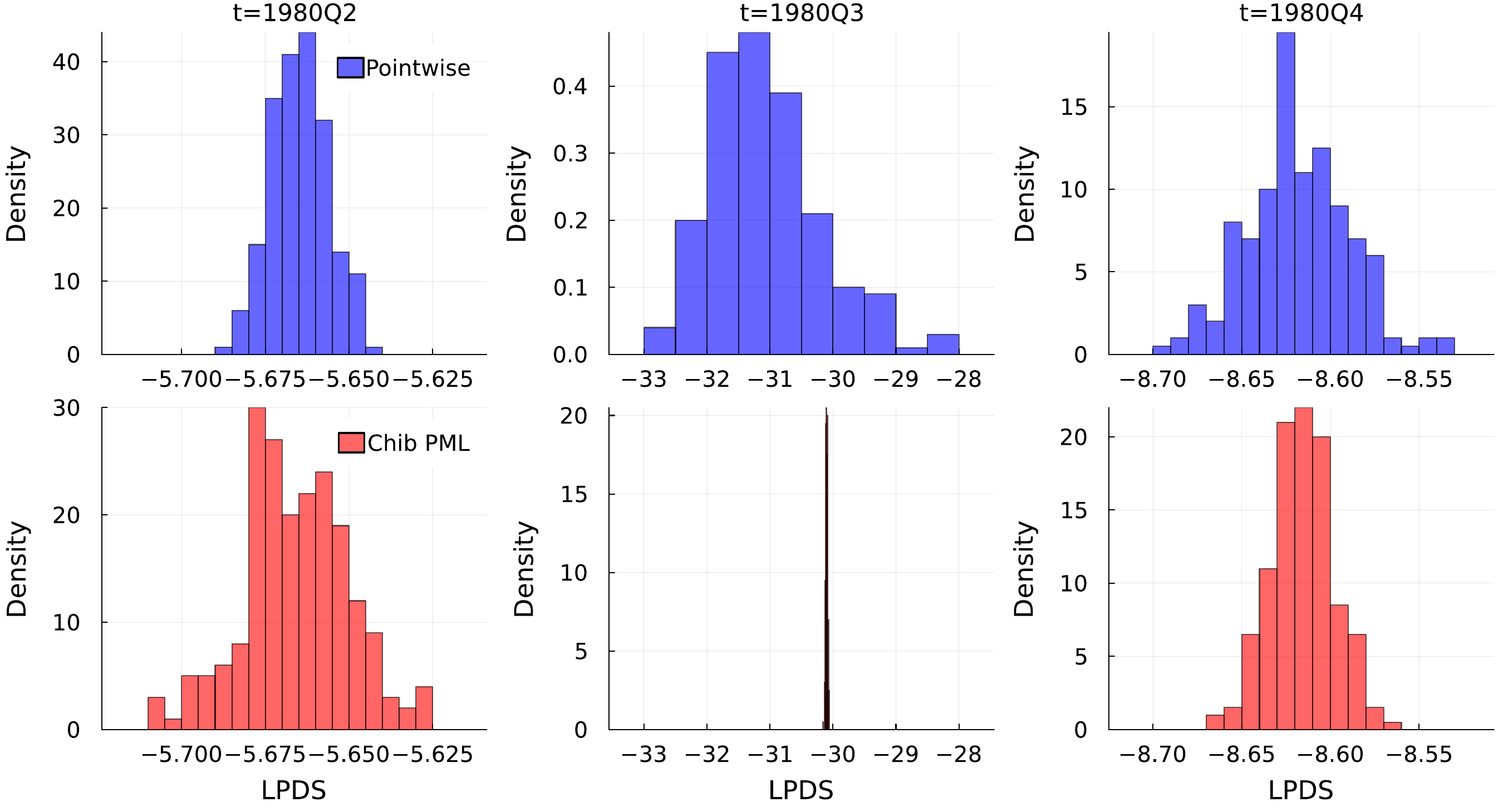}
    \caption{Bayesian VAR with normal error terms. Distribution of the estimates of the LPDS contributions for the three quarters during the Volcker shock period over $200$ runs with different seeds. Top row plots the pointwise LPDS, while the bottom row plots the PML. Each estimate is based on $20000$ Gibbs sampling draws, where 2000 were discarded as a burn-in.}
    \label{fig:CompLPDStMoreIter}
\end{figure}

A large part of the erratic behavior of the pointwise LPDS stems from a single observation, 1980Q3, a period defined by the aggressive contractionary monetary policy implemented by Paul Volcker to curb inflation. To show the effect of this outlier, Figure \ref{fig:CompLPDSt} plots the estimates of the LPDS contributions for the three time periods 1980Q2-1980Q4 across 200 repeated runs, for an overall shrinkage level of $\lambda_1=0.5$. The main difference between the pointwise LPDS estimator and the PML estimator is in 1980Q3, which is an outlier very far out in the tail giving a very low LPDS contribution. The pointwise LPDS estimator has a huge variance between runs. The PML estimator, in contrast, remains stable since it is not based on a Monte Carlo tail estimate. We can also see the bias manifest itself with the large bulk of runs clearly below the values of the PML estimator. Similar patterns are seen for different levels of $\lambda_1$, with much higher variance and a clear bias for the pointwise estimator, and the LPDS distributions are shifted horizontally.   

Figure \ref{fig:CompLPDStMoreIter} shows that an increase in the Monte Carlo sample size to $20000$ draws reduces the variability of the pointwise LPDS estimator a bit, but even with this larger posterior sample the LPDS estimate can differ by up to $4\text{-}5$ LPDS units solely on account of this single outlier; this is enough to tip the scale in favor of a suboptimal model ('strong' or 'very strong' evidence according to the modified \citet{jeffreys1961theory} scale of evidence in \citet{kass1995bayes}). Note also that our model used for illustration has only three variables, so these problems will be many times worse in the kind of large-scale models used in practice.

\subsection{Results from a model with $t$-distributed errors}\label{subsec:t-errors}

Given the instability of the pointwise LPDS estimator, which comes from the Monte Carlo approximation in the extreme tail probabilities, it is interesting to investigate whether a more heavy-tailed $t$-distributed error term in the VAR leads to a more stable estimator. 

We fix $\lambda_1$, $\lambda_2$ and $\lambda_3$ as above and use $v=5$ degrees of freedom in the $t$-distribution for illustration. The $t$-distribution is modeled as a scale-mixture of normal distributions, and the Gibbs sampler for the Gaussian BVAR is straightforwardly extended with an updating step for the augmented time-varying scales \citep{geweke1993bayesian}.

\begin{figure}
    \centering
    \includegraphics[width=0.9\linewidth]{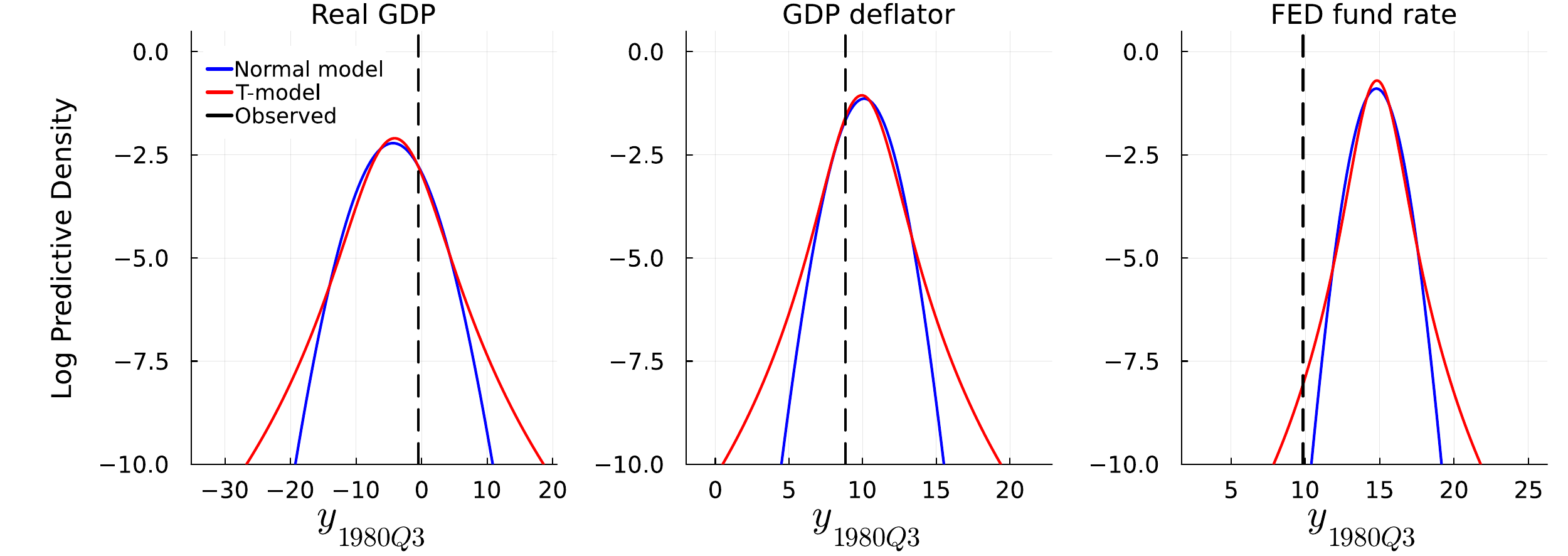}
    \caption{Marginal log predictive densities for the normal and $t$-model in 1980Q3.}
    \label{fig:marginals_t}
\end{figure}

\begin{figure}
    \centering
    \includegraphics[width=1\linewidth]{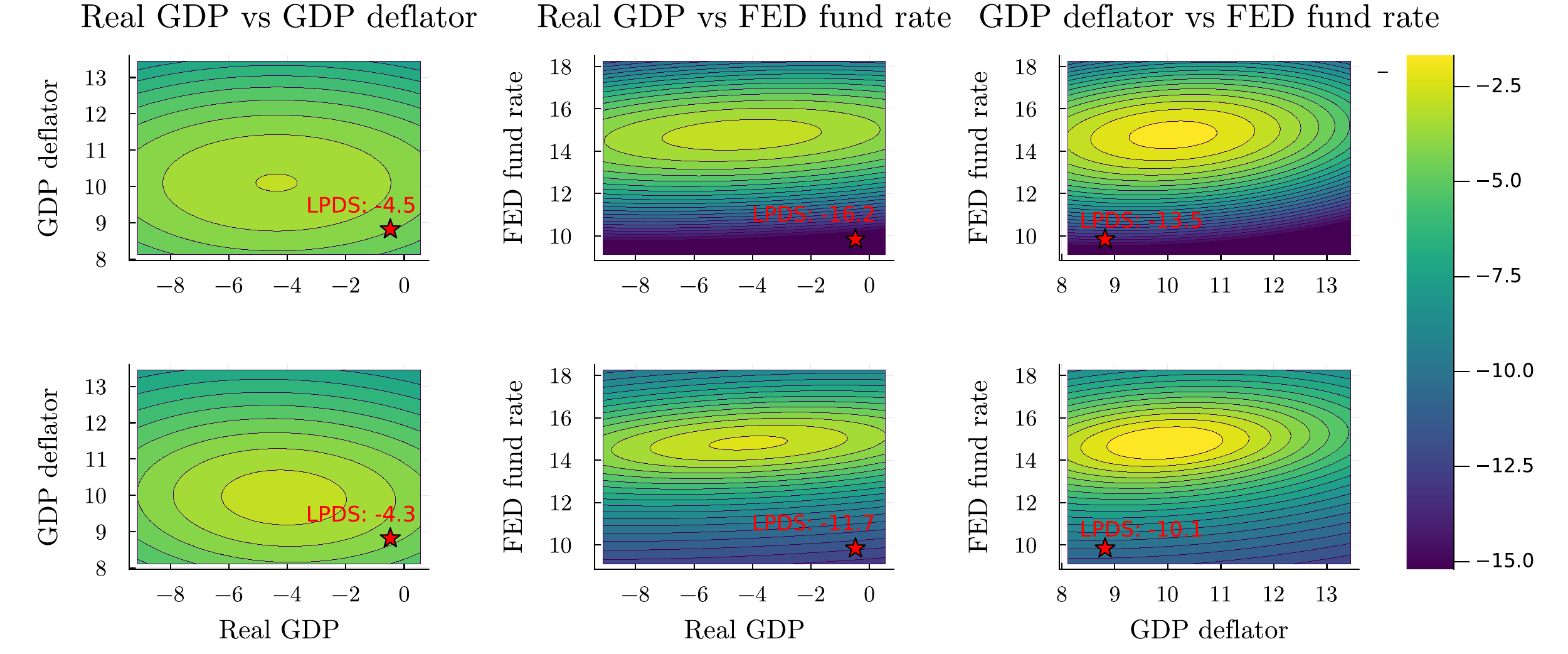}
    \caption{Bivariate log predictive distributions for 1980Q3 from the normal (top row) and $t$-model (bottom row), marginalized with respect to the third time series.}
    \label{fig:joint_t}
\end{figure}

Figure \ref{fig:marginals_t} compares the marginal predictive densities for the normal model (blue) and the Student-$t$ model (red) for 1980Q3, and the realized outlier (dashed vertical line). It is clear that the normal model fails to capture the unexpected lowering of the FED funds rate. The normal model assigns an extremely low density at the outlier, which causes the instability; in contrast, the model with $t$-distributed errors assigns enough probability mass in the tail for the evaluation to be stable; the regression example in Section 4.3 in \citet{vehtari2017practical} makes a related observation where a $t$-distribution is used to make the ELPD-LOO estimator more robust to outliers by making the full posterior a better importance sampling approximation to the desired leave-one-out posterior. Figure \ref{fig:joint_t} does a similar exercise, but plots contours of pairwise log joint predictive densities during the Volcker shock. Again, it is clear that the observation is a massive outlier in the normal model, driven by the unexpected lowering of the FED funds rate. Note, however, that in order to be able to formally show that the $t$-model is better than the normal model we still need to compute the LPDS for the normal model, so a reliable estimate of the LPDS is clearly needed also for models with outliers.

\begin{figure}
    \centering
    \includegraphics[width=0.8\linewidth]{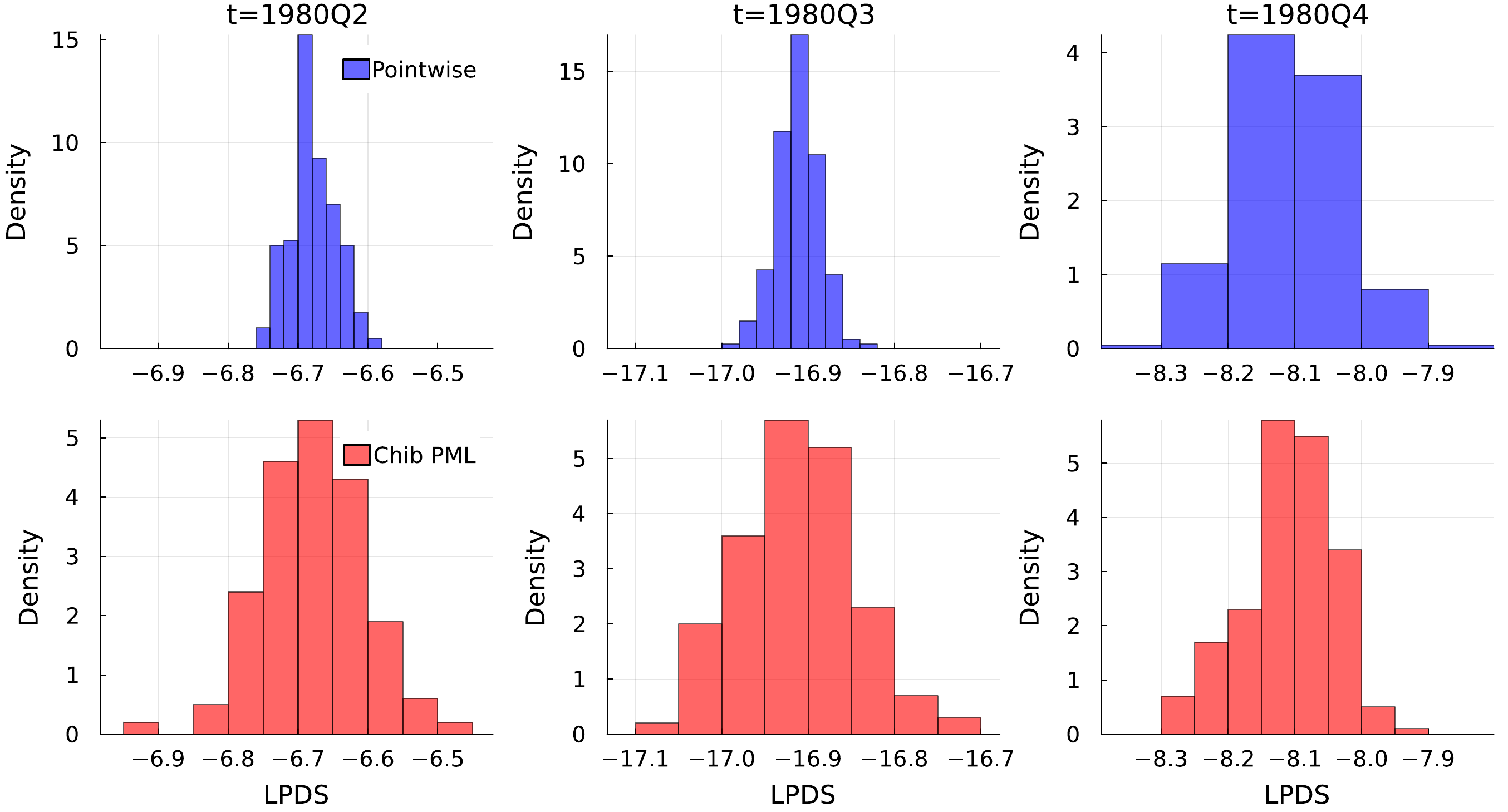}
      \caption{Bayesian VAR with $t$-distributed error terms. Distribution of the estimates of the LPDS contributions for the three quarters during the Volcker shock period over $200$ runs with different seeds. Top row plots the pointwise LPDS, while the bottom row plots the PML. Each estimate is based on $5000$ Gibbs sampling draws, where 1000 were discarded as a burn-in.}
    \label{fig:BVARTLPDSContr}
\end{figure}

\begin{figure}
    \centering
    \includegraphics[width=0.9\linewidth]{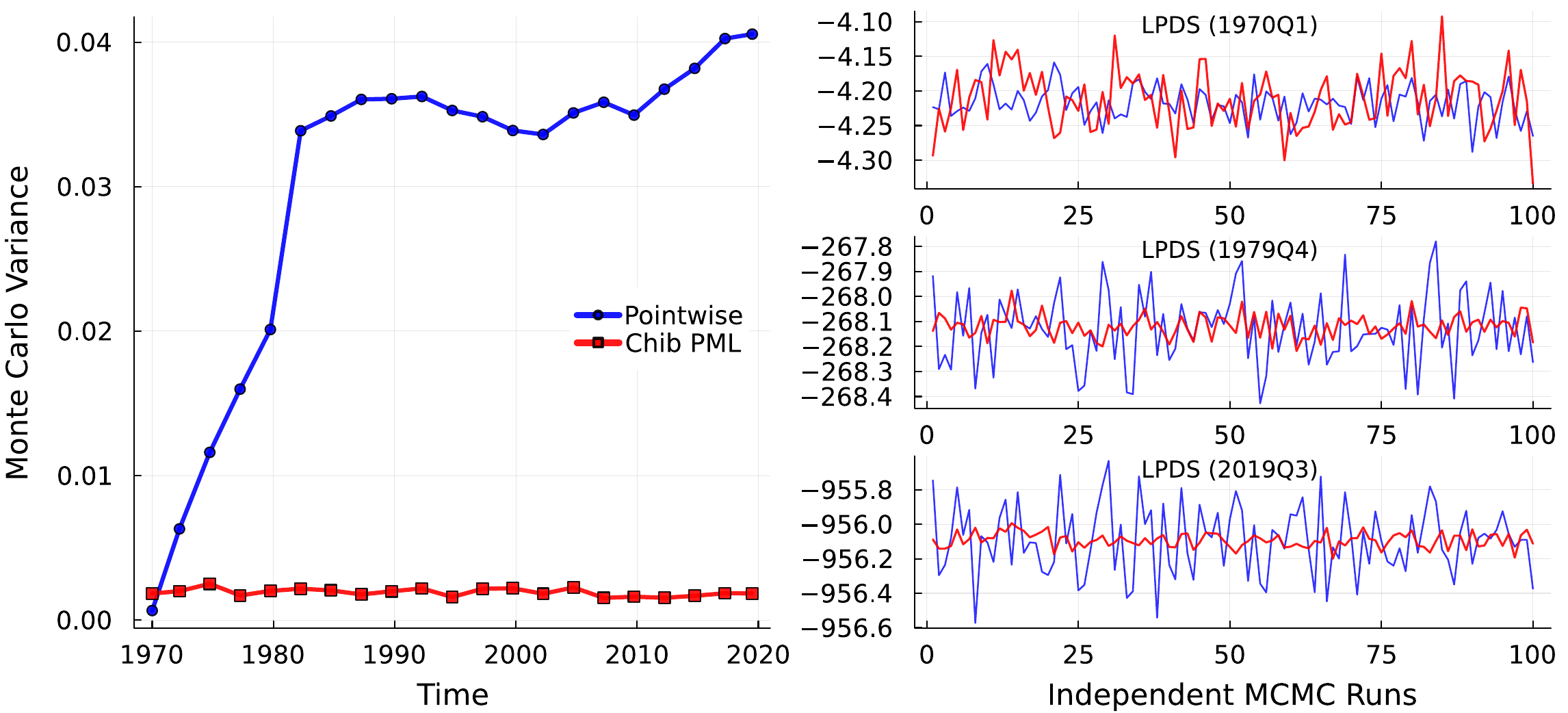}
    \caption{Comparison of the Monte Carlo variance for the cumulative LPDS using the pointwise estimator versus the Chib PML approach. \textit{Left:} The variance of the cumulative LPDS over the evaluation period 1970Q1 to 2019Q4 (in steps of 10 quarters), calculated across 100 independent MCMC runs. \textit{Right:} The realized cumulative LPDS estimates up to three selected quarters (1970Q1, 1979Q4, and 2019Q3) across the 100 independent runs.}
    \label{fig:ComparEstimVariance}
\end{figure}

Figure \ref{fig:BVARTLPDSContr} shows that when there are no outliers, Chib PML has a higher variance for individual LPDS contributions than the pointwise method. This is expected since the PML approach is based on the difference of two estimates with \textit{independent} Monte Carlo errors, so the variance of the PML is the sum of the two Monte Carlo variances. In contrast, the pointwise approach has only one source of Monte Carlo error for a \emph{single} LPDS contribution. The picture is quite different for longer evaluation periods, as shown in Figure \ref{fig:ComparEstimVariance}; this figure shows the Monte Carlo variance of the cumulative LPDS over time (left plot) and the cumulative LPDS estimates up to three selected quarters across 100 independent MCMC runs (right plot). As seen in the left panel, the Monte Carlo variance of the pointwise estimator accumulates additively at each evaluation step, causing the total uncertainty to grow substantially as the evaluation window expands. The variance of the PML estimator, on the other hand, remains completely flat and unaffected by the length of the evaluation period. Since the total LPDS in the PML framework is computed directly as the difference between two marginal likelihoods, one for the full dataset and one for the initial training set, it effectively bypasses the compounding Monte Carlo errors in the pointwise estimator.

\begin{figure}[htbp]
    \centering
    
    \begin{minipage}[c]{0.48\textwidth}
        \centering
        \includegraphics[width=\linewidth]{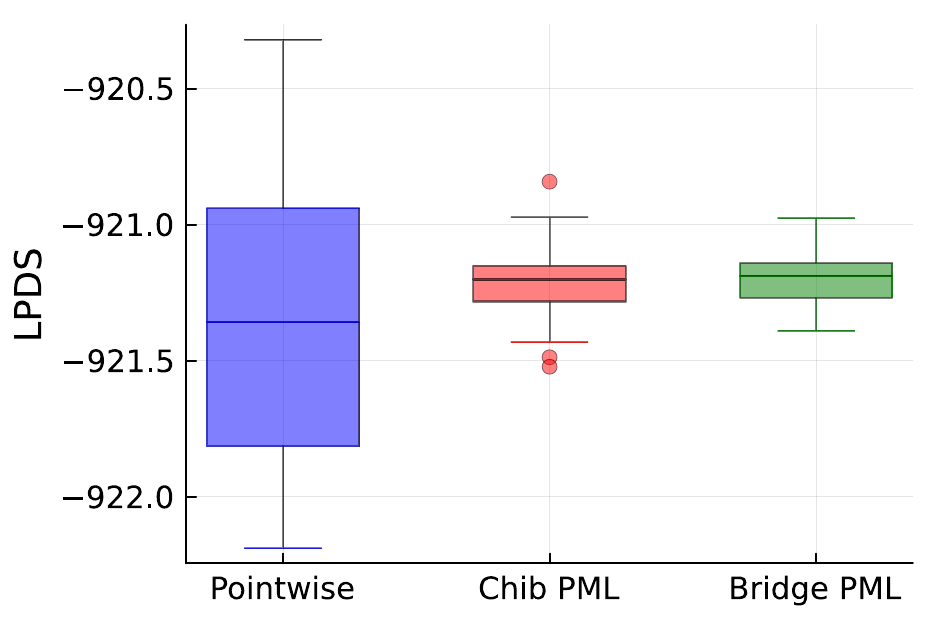}
    \end{minipage}
    \hfill
    \begin{minipage}[c]{0.48\textwidth}
    \vspace{-1cm}
        \centering
        \begin{tabular}{l c c r}
            \toprule
            \textbf{Method} & \textbf{Mean} & \textbf{SD} & \textbf{Time (s)} \\
            \midrule
            Pointwise  & $-921.34$ & $0.54$ & $27.0$ \\
            Chib PML   & $-921.22$ & $0.14$ & $2.5$ \\
            Bridge PML & $-921.20$ & $0.09$ & $470.5$ \\
            \bottomrule
        \end{tabular}
    \end{minipage}
    
    \vspace{4mm}
    
    \caption{Comparison of Total LPDS from 1970Q1 to 2019Q4 and computational times across 50 independent runs (2000 MCMC draws per run). \textit{Left:} Boxplot illustrating the variability of the LPDS estimates. \textit{Right:} Summary statistics of the LPDS and computational time.}
    \label{fig:lpdsSummary}
\end{figure}

Since Chib's method for computing the marginal likelihood is only available when we can use Gibbs sampling, and since we need to handle a quite large set of latent scales in the student-t model, we compare the results with the PML estimator based on bridge sampling from the Hamiltonian Monte Carlo output. There are at least two advantages from this: (i) it makes the PML estimator more general since it does not require Gibbs sampling, (ii) it eliminates the need for latent variables and thus heavily reduces the number of parameters, which could lead to efficiency gains. We use the same setup for the prior and the model structure as before, and we evaluate the LPDS between 1970Q1 and 2019Q4. Each estimator is based on $2000$ MCMC samples, where $500$ are discarded as burn-in. To assess the stability of the estimators, we repeated the experiment with $50$ random MCMC seeds. From Figure \ref{fig:lpdsSummary} we see that the PML estimators are very efficient (in terms of variance) compared to the pointwise estimator. Chib PML is the clear winner in terms of computational efficiency. Since bridge PML relies on HMC sampling from \texttt{Turing.jl}, and the number of parameters is fairly large, it is very slow in this example. Bridge sampling itself takes only a tiny fraction of the computational time, and does not contribute to the slow performance. 

\section{Model comparison in multiple linear regression with outliers}\label{sec:linreg_application}

For cross-sectional data, Leave-One-Out Cross-Validation (LOO-CV) is the recommended model comparison quantity in probabilistic programming frameworks such as Stan \citep{carpenter2017stan} and Turing.jl \citep{ge2018turing}. The LOO-CV is computed from a single HMC run on all the observations with an importance sampling correction for each test fold using Pareto Smoothed Importance Sampling (PSIS) \citep{vehtari2017practical}. PSIS is efficient but can fail for high-leverage outliers, as indicated by a Pareto shape parameter $k > 0.7$. 

When the PSIS diagnostic flags an observation, the standard recommendation is to perform exact HMC/MCMC refitting for that specific observation/fold and compute the pointwise LPDS estimate of the predictive density \citep{loopackage, vehtari2017practical}. However, as demonstrated in Section \ref{sec:theory}, this fallback estimator is biased and highly variable when the flagged observation is an outlier, potentially leading to erroneous model selection.
We propose to instead use the trimmed PML estimator in Section \ref{sec:trimmedPML} to estimate the LPDS contributions of the flagged observations, for example using bridge sampling.

We use a dataset on Scottish Hill races \citep{atkinson1986comment} from the MASS package in R \citep{massbook} to illustrate the gain from handling flagged observations with the PML instead of the pointwise estimator. The dataset contains $n=35$ observations with the dependent variable \textit{finishing time in minutes}, explained by the covariates $x_1\equiv$\textit{distance in miles}, and $x_2\equiv$ \textit{height in ft}. The data set is known to contain high leverage outliers. A multiple linear regression model with an intercept and two covariates is used to model the finishing times. The dataset is standardized before the analysis. 

As an illustration, we compare two competing prior specifications using the conjugate Normal-Inverse-Gamma distribution, $\sigma^2 \sim \mathcal{IG}(\nu_0/2, S_0/2)$ and $\v\beta \mid \sigma^2 \sim \mathcal{N}(\bm{0}, \sigma^2 \v V_0)$. Using a conjugate prior allows us to compute the exact marginal likelihoods and the true LOO-CV scores analytically, providing a ground truth for the evaluation. To provide a weakly informative baseline for the error variance, we set $\nu_0 = 5$ and $S_0 = \hat{\sigma}^2_{\mathrm{ols}}(\nu_0 - 2)$, where $\hat{\sigma}^2_{\mathrm{ols}}$ is the unbiased OLS residual variance of $34$ of the observations, excluding the \emph{Knock Hill} observation, which is considered the worst outlier. The two models differ only in the prior covariance of the regression coefficients:
\begin{itemize}
    \item $M_1$ - \textbf{Mild shrinkage:} A weakly informative prior with $\v V_0 = 2 \v I$.\vspace{-0.2cm}
    \item $M_2$ - \textbf{No shrinkage:} A vague prior with $\v V_0 = 500 \v I$.
\end{itemize}

We fit both models using Hamiltonian Monte Carlo (HMC) via Turing.jl, with $1000$ posterior samples per run. We evaluate the models using the Pareto smoothing importance sampling \citep{vehtari2024pareto} provided in \texttt{PSIS.jl}. During the $200$ independent HMC runs, on average (roughly) $2.4$ out of the $35$ observations are flagged as outliers and are re-evaluated using both the pointwise and the PML estimator of the LPDS contribution.

\begin{figure}
     \centering
     \includegraphics[width=0.8\linewidth]{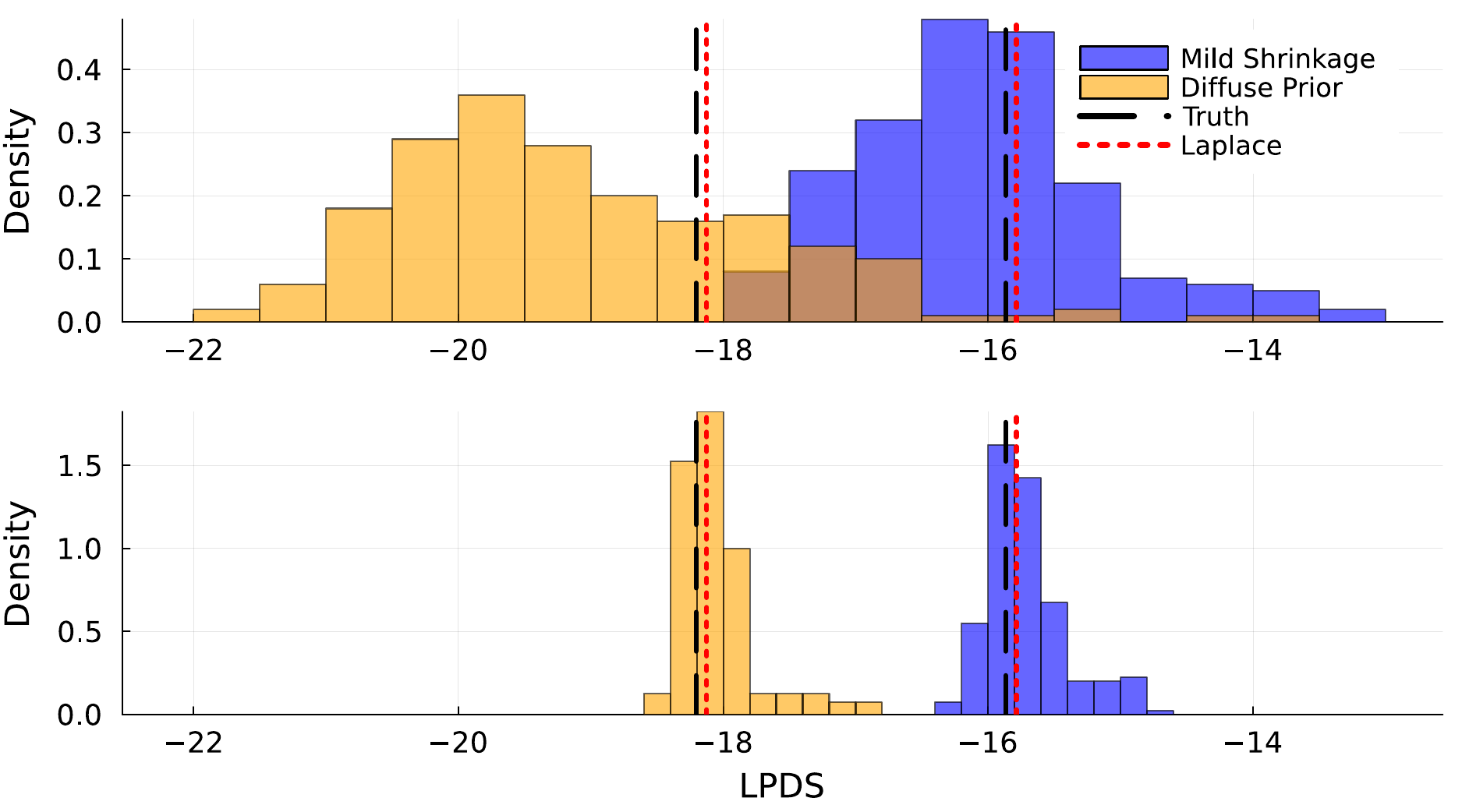}
     \caption{Distribution of the estimated total LOO-CV scores for two competing models across $400$ independent HMC runs with two different estimators for flagged observations: the pointwise LPDS in the top panel and the PML estimator via bridge sampling in the bottom panel. Dashed lines represent the true analytical values.}
     \label{fig:simulated_model_selection}
 \end{figure}

Figure \ref{fig:simulated_model_selection} shows the distribution of the total LOO-CV scores for both models across $400$ independent HMC runs. In the top panel, the outliers are evaluated using the pointwise LPDS fallback. The bias and the associated Monte Carlo variance cause the distributions of the two models to overlap; different seeds for the HMC can therefore lead to different preferred models.

The bottom panel of Figure \ref{fig:simulated_model_selection} shows the results when the LPDS contributions of flagged observations are re-estimated using the PML estimator via split-chain bridge sampling. The Monte Carlo variance is substantially reduced, and the two distributions are clearly separated, allowing for coherent model selection. 

 \begin{figure}
     \centering
     \includegraphics[width=0.9\linewidth]{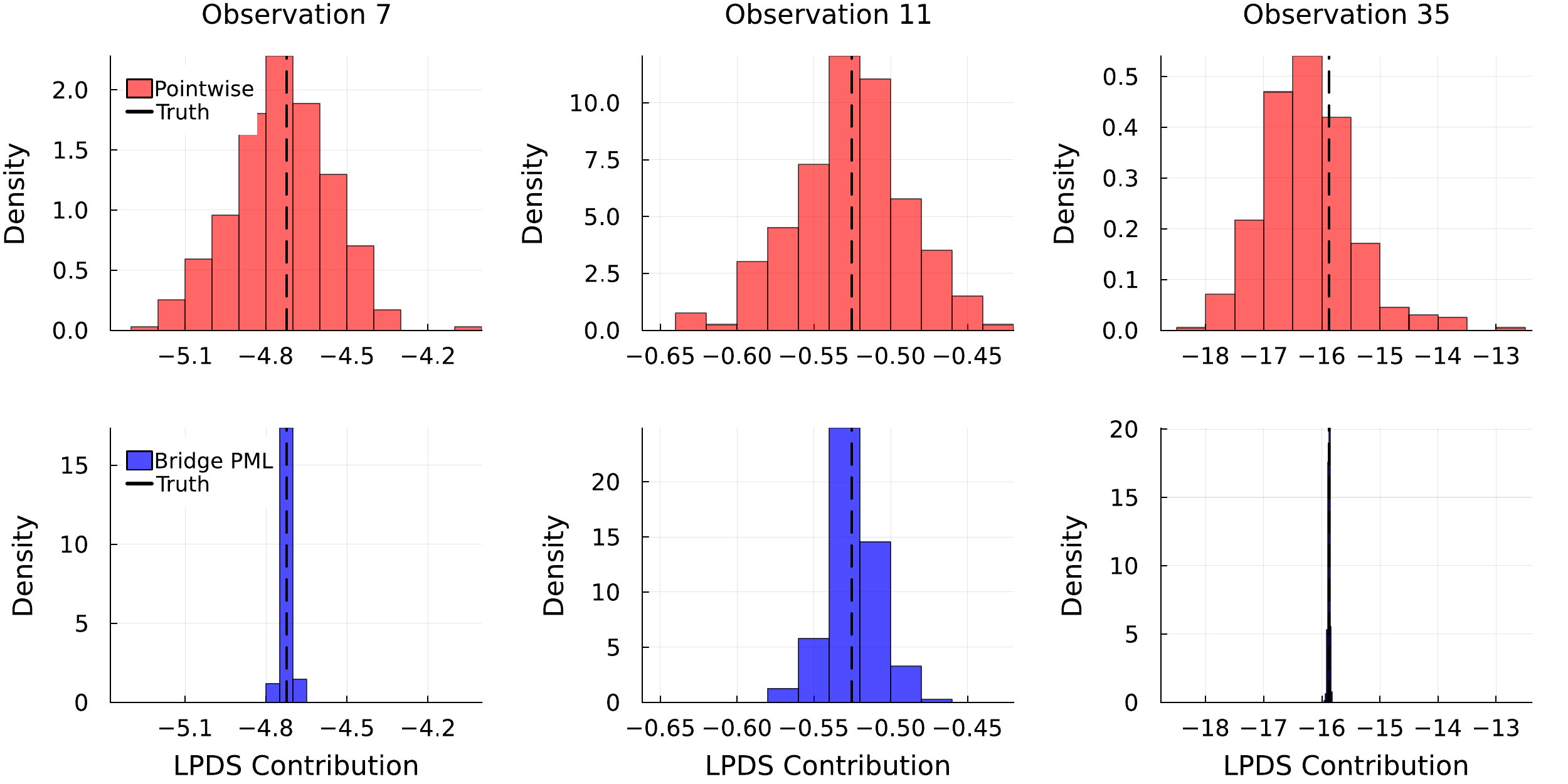}
     \caption{Distribution of $400$ independent estimates of the LPDS contributions for the three observations most often flagged by the $k$-statistic for the model with mild shrinkage. The top panel shows the results for the pointwise LPDS estimator while the second panel shows the results from the PML estimator via bridge sampling. Dashed black lines mark out the true analytical values and dotted red lines represent a Laplace approximation.}
     \label{fig:FallbakOutliers}
 \end{figure}   

 The variability of the PML estimator of LOO-CV in Figure \ref{fig:simulated_model_selection} comes mainly from the uncertainty in the observations handled by PSIS, not the flagged observations estimated by PML. To see this, Figure \ref{fig:FallbakOutliers} singles out the distribution of the estimates for the LPDS contributions of the  three observations most often flagged by the $k$-statistic for the model with mild shrinkage; the top row are estimates from the pointwise estimator, which clearly struggles with the \emph{Knock Hill} outlier (observation $35$ in the third column). The bottom panel shows that the PML estimator has a very low variance for all three outliers.

 Finally, Figure \ref{fig:simulated_model_selection} also shows the PML based on the Laplace approximation of the two marginal likelihoods as red dashed vertical lines. This estimator requires only two numerical optimization runs and is therefore very fast, and the resulting PML estimate is very close to the true LPDS. This accuracy is supported by the asymptotic analysis in \citet{gelfand1994bayesian} who show that the PML based on the Laplace approximation has an error of $O(\tilde{n}^{-2})$, where $\tilde{n}$ is the size of the training dataset, when the size of the test data is fixed. The $O(\tilde{n}^{-1})$ terms cancel since the PML is a ratio of two marginal likelihoods that share most of the observations; see \citet{tierney1986accurate}.

\section{Concluding Remarks}
The standard approach to Bayesian model comparison relies on pointwise Monte Carlo estimation of the log predictive density score (LPDS). Our article demonstrates that this estimator is highly variable and systematically biased downward in the presence of outliers, particularly when the underlying predictive distributions has thin tails. These deficiencies are severe enough to distort model comparisons and hyperparameter selection, even in low-dimensional settings. We point out that the LPDS is identical to the Partial Marginal Likelihood (PML) used in earlier literature on non-informative priors, and we show that estimating the LPDS as the difference between two marginal likelihoods provides a dramatically more stable and robust alternative.

The PML estimator is effective across different modeling domains, and it is particularly well suited for model evaluation and hyperparameter tuning in time series models. This is demonstrated for hyperparameter estimation in a steady-state vector autoregression on US macroeconomic data. We show that the PML estimator is robust to severe outliers and dramatically more stable than the commonly used estimator of the LPDS, which can yield highly variable and biased results. For cross-sectional data, we demonstrate that the PML estimator is a much better fallback estimator of the LPDS contributions of the observations flagged as problematic by the $k$-statistic in the leave-one-out cross-validation (LOO-CV) framework using Pareto smoothed importance sampling (PSIS). The PSIS with the PML as fallback method is very stable across seeds, effective in reducing bias, and retains a highly efficient computational budget by only targeting flagged outliers.

While the PML estimator successfully eliminates the local tail-evaluation problem, its overall success relies on the availability of accurate marginal likelihood estimators. In very high-dimensional models or settings with highly irregular posteriors, computing marginal likelihoods becomes increasingly difficult. However, since the bias of the pointwise estimator also deteriorates in high dimensions, evaluating the relative robustness of these estimators in highly complex settings is an important direction for future research.

\section*{Acknowledgments}
The authors were partially funded by the Swedish Research Council, grant 2025-01654. The computations were enabled by resources provided by the National Academic Infrastructure for Supercomputing in Sweden (NAISS), partially funded by the Swedish Research Council through grant agreement no. 2022-06725.

\singlespacing 

\bibliographystyle{apalike}
\bibliography{ref}

@article{atkinson1986comment,
  author  = {Atkinson, A. C.},
  title   = {Comment: {A}spects of Diagnostic Regression Analysis},
  journal = {Statistical Science},
  year    = {1986},
  volume  = {1},
  number  = {3},
  pages   = {397--402},
  doi     = {10.1214/ss/1177013622}
}

@book{bernardo1994bayesian,
  title={Bayesian theory},
  author={Bernardo, Jos{\'e} M and Smith, Adrian FM},
  volume={586},
  year={1994},
  publisher={Wiley Online Library}
}

@article{bitto2019achieving,
  title={Achieving shrinkage in a time-varying parameter model framework},
  author={Bitto, Angela and Fr{\"u}hwirth-Schnatter, Sylvia},
  journal={Journal of Econometrics},
  volume={210},
  number={1},
  pages={75--97},
  year={2019},
  publisher={Elsevier}
}

@article{carpenter2017stan,
  title={Stan: A probabilistic programming language},
  author={Carpenter, Bob and Gelman, Andrew and Hoffman, Matthew D and Lee, Daniel and Goodrich, Ben and Betancourt, Michael and Brubaker, Marcus and Guo, Jiqiang and Li, Peter and Riddell, Allen},
  journal={Journal of Statistical Software},
  volume={76},
  pages={1--32},
  year={2017}
}

@article{chib1995marginal,
  title={Marginal likelihood from the {Gibbs} output},
  author={Chib, Siddhartha},
  journal={Journal of the American Statistical Association},
  volume={90},
  number={432},
  pages={1313--1321},
  year={1995},
  publisher={Taylor \& Francis}
}

@inproceedings{ge2018turing,
  title={Turing: a language for flexible probabilistic inference},
  author={Ge, Hong and Xu, Kai and Ghahramani, Zoubin},
  booktitle={International Conference on Artificial Intelligence and Statistics (AIStats)},
  pages={1682--1690},
  year={2018},
  organization={PMLR}
}

@book{gelman2013bayesian,
  title={Bayesian Data Analysis},
  author={Gelman, A. and Carlin, J.B. and Stern, H.S. and Dunson, D.B. and Vehtari, A. and Rubin, D.B.},
  isbn={9781439898208},
  lccn={2013039507},
  series={Chapman \& Hall/CRC Texts in Statistical Science},
  url={https://books.google.se/books?id=eSHSBQAAQBAJ},
  year={2013},
  publisher={CRC Press}
}

@article{geweke1993bayesian,
  title={Bayesian treatment of the independent Student-t linear model},
   author={Geweke, John},
  journal={Journal of Applied Econometrics},
  volume={8},
  number={S1},
  pages={S19--S40},
  year={1993},
  publisher={Wiley Online Library}
}

@article{geweke2010comparing,
  title={Comparing and evaluating {Bayesian} predictive distributions of asset returns},
  author={Geweke, John and Amisano, Gianni},
  journal={International Journal of Forecasting},
  volume={26},
  number={2},
  pages={216--230},
  year={2010},
  publisher={Elsevier}
}

@article{gustafsson2023bayesian,
  title={Bayesian optimization of hyperparameters from noisy marginal likelihood estimates},
  author={Gustafsson, Oskar and Villani, Mattias and Stockhammar, P{\"a}r},
  journal={Journal of Applied Econometrics},
  volume={38},
  number={4},
  pages={577--595},
  year={2023},
  publisher={Wiley Online Library}
}

@article{kass1993bayes,
  title={Bayes factors in practice},
  author={Kass, Robert E},
  journal={Journal of the Royal Statistical Society: Series D},
  volume={42},
  number={5},
  pages={551--560},
  year={1993},
  publisher={Wiley Online Library}
}

@article{meng1996simulating,
  title={Simulating ratios of normalizing constants via a simple identity: a theoretical exploration},
  author={Meng, Xiao-Li and Wong, Wing Hung},
  journal={Statistica Sinica},
  volume = {6},
  pages={831--860},
  year={1996},
  publisher={JSTOR}
}

@article{sivula2025uncertainty,
  title={Uncertainty in {Bayesian} leave-one-out cross-validation based model comparison},
  author={Sivula, Tuomas and Magnusson, M{\aa}ns and Matamoros, Asael Alonzo and Vehtari, Aki},
  journal={Bayesian Analysis},
  volume={1},
  number={1},
  pages={1--31},
  year={2025},
  publisher={International Society for Bayesian Analysis}
}

@article{vehtari2012survey,
  author  = {Vehtari, A. and Ojanen, J.},
  title   = {A survey of {B}ayesian predictive methods for model assessment, selection and comparison},
  journal = {Statistics Surveys},
  volume  = {6},
  pages   = {142--228},
  year    = {2012}
}

@article{vehtari2017practical,
  title={Practical {Bayesian} model evaluation using leave-one-out cross-validation and WAIC},
  author={Vehtari, Aki and Gelman, Andrew and Gabry, Jonah},
  journal={Statistics and Computing},
  volume={27},
  number={5},
  pages={1413--1432},
  year={2017},
  publisher={Springer}
}

@article{vehtari2024pareto,
  title={Pareto smoothed importance sampling},
  author={Vehtari, Aki and Simpson, Daniel and Gelman, Andrew and Yao, Yuling and Gabry, Jonah},
  journal={Journal of Machine Learning Research},
  volume={25},
  number={72},
  pages={1--58},
  year={2024}
}

@article{villani2009steady,
  title={Steady-state priors for vector autoregressions},
  author={Villani, Mattias},
  journal={Journal of Applied Econometrics},
  volume={24},
  number={4},
  pages={630--650},
  year={2009},
  publisher={Wiley Online Library}
}

@article{chib2026testing,
  title={Testing for Endogeneity: A Moment-Based {Bayesian} Approach},
  author={Chib, Siddhartha and Shin, Minchul and Simoni, Anna},
  journal={arXiv preprint arXiv:2603.07780},
  year={2026}
}

@article{chib2025structural,
  title={Structural breaks, model uncertainty and factor selection},
  author={Chib, Siddhartha and Smith, Simon C},
  journal={Journal of Econometrics},
  pages={106067},
  year={2025},
  publisher={Elsevier}
}

@article{o1995fractional,
  title={Fractional Bayes factors for model comparison},
  author={O'Hagan, Anthony},
  journal={Journal of the Royal Statistical Society: Series B},
  volume={57},
  number={1},
  pages={99--118},
  year={1995},
  publisher={Wiley Online Library}
}

@article{berger1996intrinsic,
  title={The intrinsic {Bayes} factor for model selection and prediction},
  author={Berger, James O and Pericchi, Luis R},
  journal={Journal of the American Statistical Association},
  volume={91},
  number={433},
  pages={109--122},
  year={1996},
  publisher={Taylor \& Francis}
}

@article{gelfand1994bayesian,
  title={Bayesian model choice: asymptotics and exact calculations},
  author={Gelfand, Alan E and Dey, Dipak K},
  journal={Journal of the Royal Statistical Society: Series B},
  volume={56},
  number={3},
  pages={501--514},
  year={1994},
  publisher={Wiley Online Library}
}

@article{fong2020marginal,
  title={On the marginal likelihood and cross-validation},
  author={Fong, Edwin and Holmes, Chris C},
  journal={Biometrika},
  volume={107},
  number={2},
  pages={489--496},
  year={2020},
  publisher={Oxford University Press}
}

@article{dawid1984present,
  title={Present position and potential developments: Some personal views statistical theory the prequential approach},
  author={Dawid, A Philip},
  journal={Journal of the Royal Statistical Society: Series A},
  volume={147},
  number={2},
  pages={278--290},
  year={1984},
  publisher={Wiley Online Library}
}

@article{aitkin1991posterior,
  title={{Posterior Bayes factors}},
  author={Aitkin, Murray},
  journal={Journal of the Royal Statistical Society: Series B},
  volume={53},
  number={1},
  pages={111--128},
  year={1991},
  publisher={Oxford University Press}
}

@article{pitt2012some,
  title={On some properties of {Markov chain Monte Carlo} simulation methods based on the particle filter},
  author={Pitt, Michael K and dos Santos Silva, Ralph and Giordani, Paolo and Kohn, Robert},
  journal={Journal of Econometrics},
  volume={171},
  number={2},
  pages={134--151},
  year={2012},
  publisher={Elsevier}
}

@article{berard2014lognormal,
  title={A lognormal central limit theorem for particle approximations of normalizing constants},
  author={B{\'e}rard, Jean and Del Moral, Pierre and Doucet, Arnaud},
  journal = {Electronic Journal of Probability},
  volume = {19},
  pages = {1--28},
  year={2014}
}

@article{carriero2012forecasting,
  title={Forecasting government bond yields with large {Bayesian} vector autoregressions},
  author={Carriero, Andrea and Kapetanios, George and Marcellino, Massimiliano},
  journal={Journal of Banking \& Finance},
  volume={36},
  number={7},
  pages={2026--2047},
  year={2012},
  publisher={Elsevier}
}

@article{giannone2015prior,
  title={Prior selection for vector autoregressions},
  author={Giannone, Domenico and Lenza, Michele and Primiceri, Giorgio E},
  journal={Review of Economics and Statistics},
  volume={97},
  number={2},
  pages={436--451},
  year={2015},
  publisher={The MIT Press}
}

@article{doan1984forecasting,
  title={Forecasting and conditional projection using realistic prior distributions},
  author={Doan, Thomas and Litterman, Robert and Sims, Christopher},
  journal={Econometric Reviews},
  volume={3},
  number={1},
  pages={1--100},
  year={1984},
  publisher={Taylor \& Francis}
}

@Misc{loopackage,
    title = {loo: Efficient leave-one-out cross-validation and {WAIC} for
      {Bayesian} models},
    author = {Vehtari, Aki and Gabry, Jonah and Magnusson, Måns and Yao, Yuling  and Bürkner, Paul-Christian and Paananen, Topi and Gelman, Andrew},
    year = {2026},
    note = {R package version 2.10.1},
    url = {https://mc-stan.org/loo/}
  }

@Book{massbook,
    entry = {Book},
    title = {Modern Applied Statistics with S},
    author = {Ripley, Brian D. and Venables, William N.},
    publisher = {Springer},
    edition = {Fourth},
    address = {New York},
    year = {2002},
    url = {https://www.stats.ox.ac.uk/pub/MASS4/},
  }

@article{kass1995bayes,
  title={Bayes factors},
  author={Kass, Robert E and Raftery, Adrian E},
  journal={Journal of the American Statistical Association},
  volume={90},
  number={430},
  pages={773--795},
  year={1995},
  publisher={Taylor \& Francis}
}

@book{jeffreys1961theory,
  title={Theory of probability},
  author={Jeffreys, Harold},
  year={1961},
  publisher={Oxford University Press}
}

@article{geisser1979predictive,
  title={A predictive approach to model selection},
  author={Geisser, Seymour and Eddy, William F},
  journal={Journal of the American Statistical Association},
  volume={74},
  number={365},
  pages={153--160},
  year={1979},
  publisher={Taylor \& Francis}
}

@article{tierney1986accurate,
  title={Accurate approximations for posterior moments and marginal densities},
  author={Tierney, Luke and Kadane, Joseph B},
  journal={Journal of the American Statistical Association},
  volume={81},
  number={393},
  pages={82--86},
  year={1986},
  publisher={Taylor \& Francis}
}

\doublespacing 

\newpage 


\title{ Supplement to \\ Log predictive density score estimation in the presence of outliers}
\if0\blind
{
\author{Oskar Gustafsson$^{a}$\thanks{Corresponding author: oskar.gustafsson@stat.su.se. 
\noindent $^a$Department of Statistics, Stockholm University.} and Mattias Villani$^{a}$}
}\fi
\date{}
\maketitle

\renewcommand{\theequation}{S\arabic{equation}}
\renewcommand{\thesection}{S\arabic{section}}
\renewcommand{\theproposition}{S\arabic{proposition}}
\renewcommand{\theassumption}{S\arabic{assumption}}

\renewcommand{\thethm}{S\arabic{thm}}

\renewcommand{\thelemma}{S\arabic{lemma}}
\renewcommand{\thealgocf}{S\arabic{algocf}}
\renewcommand{\thefigure}{S\arabic{figure}}
\renewcommand{\thetable}{S\arabic{table}}
\renewcommand{\thepage}{S\arabic{page}}
\renewcommand{\thetable}{S\arabic{table}}
\renewcommand{\thepage}{S\arabic{page}}
\setcounter{page}{1}
\setcounter{section}{0}
\setcounter{equation}{0}
\setcounter{algocf}{0}
\setcounter{lemma}{0}
\setcounter{assumption}{0}
\setcounter{table}{0}
\setcounter{figure}{0}
\setcounter{thm}{0}
\numberwithin{equation}{section}

\begin{abstract}
    This Supplement contains proofs and additional results for the paper '\emph{Log predictive density score estimation in the presence of outliers}'.
\end{abstract}

\BlankLine

\section{Bias and variance in the multivariate normal model with known variance - proof and assessment of accuracy} \label{sec:deriveBias}

The section derives approximate expressions for the mean, variance, and skewness of the pointwise LPDS estimator from properties of the non-central $\chi^2$ distribution. The accuracy of the approximation is evaluated in an illustrative example by simulation.

The non-central $\chi^2$ is defined as a sum of independent normal random variables with variance 1. Since $\mu \sim N(\hat{\mu}_t,c\Sigma)$, we have $y_{t+1} - \mu^{(m)} \sim N(y_{t+1} - \hat{\mu}_t,c\Sigma)$. To make $Q^{(m)}$ above have the data covariance matrix $\Sigma$, we need to scale $y_{t+1} - \mu^{(m)}$ by $1/\sqrt{c}$.  
The standardized distance $Z = Q^{(m)}/c$ now follows a non-central chi-squared distribution, $Z \sim \chi^2_K(\lambda)$, with $K$ degrees of freedom and a non-centrality parameter:
\begin{equation}
    \lambda = \frac{1}{c}D^2,
\end{equation} where $D^2= (y_{t+1} - \hat{\mu}_t)^T \Sigma^{-1} (y_{t+1} - \hat{\mu}_t)$ is the squared Mahalanobis distance of the new data point $y_{t}$ from the posterior mean. As $y_{t+1}$ moves into the tails, $\lambda \to \infty$ and $p \to 0$ lead to an increasing Taylor bias. We note again that the above is for \textbf{one} of the $m$ posterior draws, i.e., $\hat{p}_1$.

Let $C = (2\pi)^{-K/2} |\Sigma|^{-1/2}$ denote the normalizing constant of the multivariate normal distribution, such that the ``naive'' estimator is exactly $\widehat{p} = C \exp(-\frac{c}{2} Z)$. The moments of this estimator can be derived using the \textit{Moment Generating Function} (MGF) of the non-central chi-squared distribution, $M_Z(t) = \mathbb{E}[e^{tZ}] = (1 - 2t)^{-K/2} \exp\left( \frac{\lambda t}{1 - 2t} \right)$. 

The exact first and second moments of the density estimator are:
\begin{align}
    \mathbb{E}[\widehat{p}] &= C \cdot M_Z(-c/2) = C (1+c)^{-K/2} \exp\left( -\frac{\lambda c}{2(1+c)} \right) \\
    \mathbb{E}[\widehat{p}^2] &= C^2 \cdot M_Z(-c) = C^2 (1+2c)^{-K/2} \exp\left( -\frac{\lambda c}{1+2c} \right).
\end{align}

Note that the first moment perfectly recovers the exact analytical predictive density, $p(y_{t+1} \mid y_{1:t}) = \mathcal{N}(y_{t+1} \mid \hat{\mu}_t, (1+c)\Sigma)$, which confirms the unbiasedness of the naive Monte Carlo estimator. However, it is the second moment, $\mathbb{E}[\widehat{p}^2]$, that dictates the variance and consequently the Taylor bias.

The variance $\text{Var}(\widehat{p}) = \mathbb{E}[\widehat{p}^2] - (\mathbb{E}[\widehat{p}])^2$ determines the magnitude of the Taylor bias. Conveniently, the constant $C^2$ factors out in both the numerator and the denominator, leaving the relative variance unaffected by the normalization:

\begin{equation}
\begin{aligned}
    \text{Bias} &\approx - \frac{\text{Var}(\widehat{p})}{2 (\mathbb{E}[\widehat{p}])^2} \\
    &= - \frac{1}{2} \left( \frac{\mathbb{E}[\widehat{p}^2] - (\mathbb{E}[\widehat{p}])^2}{(\mathbb{E}[\widehat{p}])^2} \right) \\
    &= - \frac{1}{2} \left( \frac{\mathbb{E}[\widehat{p}^2]}{(\mathbb{E}[\widehat{p}])^2} - 1 \right) \\
    &= - \frac{1}{2} \left( \frac{C^2 \cdot M_Z(-c)}{C^2 \cdot [M_Z(-c/2)]^2} - 1 \right) \\
    &= - \frac{1}{2} \left( \frac{M_Z(-c)}{[M_Z(-c/2)]^2} - 1 \right) \\
    &= - \frac{1}{2} \left[ \frac{(1+2c)^{-K/2} \exp\left( -\frac{\lambda c}{1+2c} \right)}{(1+c)^{-K} \exp\left( -\frac{\lambda c}{1+c} \right)} - 1 \right] \\
    &= - \frac{1}{2} \left[ \left( 1 + \frac{c^2}{1+2c} \right)^{K/2} \exp\left( \frac{\lambda c^2}{(1+c)(1+2c)} \right) - 1 \right].
\end{aligned}
\end{equation}

Let $\mu_k = \mathbb{E}[\widehat{p}^k]$ denote the raw moments of the predictive density estimator. Following the moment-generating function of the underlying quadratic form, the third (first and second are given above) raw moment is:
\begin{align}
    \mu_3 &= C^3(1+3c)^{-K/2} \exp\left(-\frac{3\lambda c}{2(1+3c)}\right)
\end{align}

The third centered moment, $\overline{m}_3 = \mathbb{E}[(\widehat{p} - \mathbb{E}[\widehat{p}])^3]$, expands to $\overline{m}_3 = \mu_3 - 3\mu_2\mu_1 + 2\mu_1^3$. Substituting the raw moments yields the exact third centered moment:
\begin{equation}
    \overline{m}_3 = C^3 \left[ \frac{e^{-\frac{3\lambda c}{2(1+3c)}}}{(1+3c)^{K/2}} - 3\frac{e^{-\lambda c \left( \frac{1}{1+2c} + \frac{1}{2(1+c)} \right)}}{((1+2c)(1+c))^{K/2}} + 2\frac{e^{-\frac{3\lambda c}{2(1+c)}}}{(1+c)^{3K/2}} \right]
\end{equation}

We use a small simulation study to illustrate that the bias approximation \eqref{eq:bias} is accurate and to study numerically how the bias increases with the dimensionality, $K$. The mean of the data-generating process is set to $\mu_{true} = \bm{0}$. For the univariate model, we assume a known variance of $\sigma^2=1$. For the $K=20$ dimensional model, the variables are correlated via a Toeplitz covariance matrix $\Sigma$ with $\rho=0.5$ and unit diagonal elements. To mirror the exact conditions under which our theoretical bias was derived, we evaluate the estimator using a fixed posterior distribution centered at $\hat{\mu} = \mu_{true}$ and we set $c_t = 0.25$. We draw $M=50$ independent draws from the posterior to compute the estimate $\log \hat{p}_t$. This is replicated $10000$ times to obtain robust measures of the estimator's bias and standard deviation. Note that the small Monte Carlo sample size $M=50$ is used for illustration since this toy model is chosen to be very simplistic for illustration purposes.

\begin{figure}
    \centering
    \includegraphics[width=0.9\linewidth]{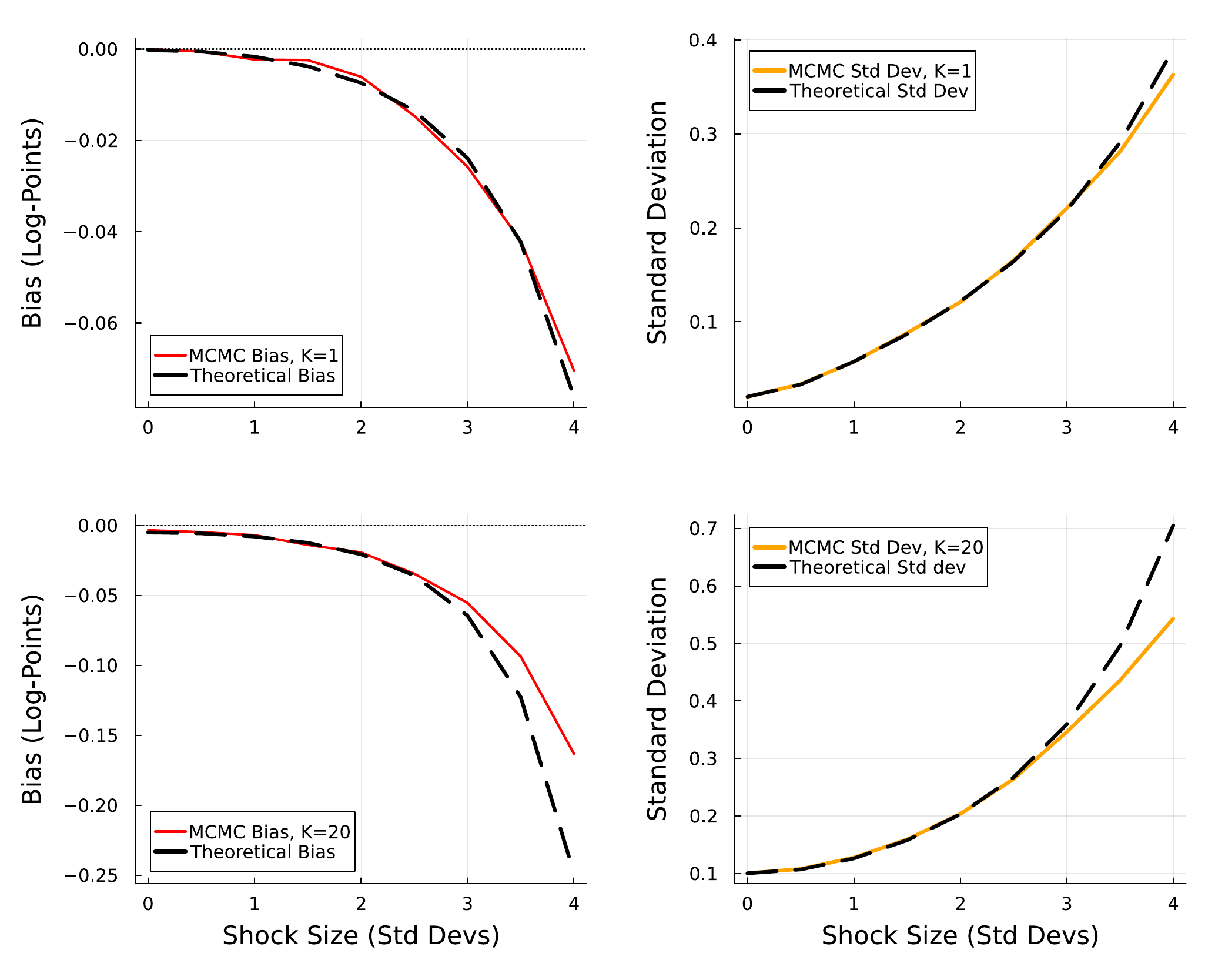}
    \caption{Comparison of the estimation bias (left) and MCMC standard deviation (right) for $K=1$ and $K=20$. The theoretical Taylor approximation (dashed black line) closely tracks the empirical MCMC bias (red line).}
    \label{fig:normalDimCompare}
\end{figure}

To explore the effect of an outlier, we introduce a structural shock by shifting the target observation $y_t$ away from the mean in increments of standard deviations (in the multivariate model, only the first variable is shifted), hence increasing the Mahalanobis distance $D_t$. Figure \ref{fig:normalDimCompare} displays the bias and standard deviation as a function of the shock size for a univariate ($K=1$) observation (top row) and a higher-dimensional ($K=20$) model (bottom row). The graphs show that both the MCMC bias and standard errors closely follow our theoretical trajectories, at least up to three standard deviations; the divergence beyond that point is expected since the bias is approximated by a local second-order Taylor expansion. The bias standard error increases exponentially with the severity of the outlier. Both the baseline bias at zero Mahalanobis distance and the exponential growth rate increase with dimensionality $K$. 

In this toy example, we need a large outlier for the bias to be non-negligible, even in a small dataset. In realistic models with more parameters, higher complexity, and with real data, $c_t$ is much higher, and the bias can be substantial.

\end{document}